\documentclass[12pt,reqno]{amsart}

\usepackage{a4}
\usepackage{amsmath}
\usepackage{amssymb}

\usepackage{graphicx}               

\theoremstyle{plain}            
\newtheorem{theorem}{Theorem}[section]
\newtheorem*{theorem*}{Theorem}
\newtheorem*{proposition}{Proposition}
\newtheorem{lemma}[theorem]{Lemma}
\newtheorem{corollary}[theorem]{Corollary}

\theoremstyle{definition}       
\newtheorem{definition}[theorem]{Definition}

\theoremstyle{remark}           
\newtheorem{remark}[theorem]{Remark}

\numberwithin{equation}{section}

\DeclareMathOperator{\card}   {card}
\DeclareMathOperator{\dist}   {dist}
\DeclareMathOperator{\dom}    {dom}
\DeclareMathOperator{\essspec}{ess\,spec}
\DeclareMathOperator{\Hom}    {Hom}
\DeclareMathOperator{\spec}   {spec}
\DeclareMathOperator{\supp}   {supp}
\DeclareMathOperator{\vol}    {vol}

\newlength{\maxbreite}%
\newlength{\maxhoehe}%
\newlength{\maxtiefe}%

\newcommand{\stelldrueber}[3][0pt]{
  \settowidth{\maxbreite}{#3}%
  \settoheight{\maxhoehe}{#3}%
  \settodepth{\maxtiefe}{#2}%
  \addtolength{\maxhoehe}{\maxtiefe}%
  {\makebox[\maxbreite]{\raisebox{\maxhoehe}{\hspace{#1}#2}}%
  \makebox[0pt][r]{#3}}%
}

\newcommand{\overcirc}[1]       
{\stelldrueber[.45ex]{$\scriptscriptstyle\circ$}{${#1}$}}

\newcommand{\R}{\mathbb{R}}                 
\newcommand{\C}{\mathbb{C}}                 
\newcommand{\N}{\mathbb{N}}                 
\newcommand{\Z}{\mathbb{Z}}                 
\newcommand{\Sphere}{\mathbb{S}}            
\newcommand{\Torus}{\mathbb{T}}             
\newcommand{\eps}{\varepsilon}              
\newcommand{\dd}{\mathrm d}                 

\newcommand{\HS}{\mathcal H}                

\newcommand{\per}[1]{\mathcal {#1}}         
\newcommand{\Mper}{\per M}                  
\newcommand{\Meps}{{M_\eps}}                
\newcommand{\Mepsper}{{\Mper_\eps}}         

\newcommand{\pert}[2]{#1_{#2}}           
\newcommand{\Mtau}[1][{}]{\pert M {\tau #1}} 
\newcommand{\Mpertau}[1][{}]{\pert \Mper {\tau #1}} 
\newcommand{\Pert}{M^m}                     
\newcommand{\Perttau}{\pert \Pert \tau}     
\newcommand{\Pertplus}{M^{m_+}}             
\newcommand{\Pertplustau}{\pert \Pertplus \tau} 
\newcommand{\Appr}{M{}^n}                   
\newcommand{\Apprm}{M{}^{n_m}}              
\newcommand{\Apprtau}[1][{}]{\pert \Appr {\tau #1}} 
\newcommand{\Rest}{R^m}                     
\newcommand{\RestAppr}{R^{m,n}}             
\newcommand{\RestApprtau}[1][{}]{\pert \RestAppr {\tau #1}} 
\newcommand{\GnM}{\Gamma^n \! M}            
\newcommand{\GpM}{\Gamma' \! M}             
\newcommand{\pointed}[1]{\dot{#1}}          
\newcommand{\MperP}{\pointed \Mper}         
\newcommand{\hMper}{\hat \Mper}         
\newcommand{\tMpertau}{\pert {\widetilde \Mper} \tau}  

\newcommand{\Neu}{{\mathrm N}}              
\newcommand{\Dir}{{\mathrm D}}              

\newcommand{\dimEW}[2][{\lambda}]{\dim_{#1}(#2)}       
\newcommand{\dimEWD}[2][{\lambda}]{\dim_{#1}^\Dir(#2)} 
\newcommand{\dimEWN}[2][{\lambda}]{\dim_{#1}^\Neu(#2)} 
\newcommand{\dimEWND}[2][{\lambda}]{\dim_{#1}^{\Neu \Dir}(#2)} 

\newcommand{\laplacian}[1]{\Delta_{{#1}}}   
\newcommand{\laplacianD}[1]{\Delta^\Dir_{{#1}}}
\newcommand{\laplacianN}[1]{\Delta^\Neu_{{#1}}}
\newcommand{\laplacianT}[1]{\Delta^{\theta}_{{#1}}}
\newcommand{\EW}[2]{\lambda_{#1}({#2})}     

\newcommand{\EWD}[2]{\lambda^\Dir_{#1}({#2})}
\newcommand{\EWN}[2]{\lambda^\Neu_{#1}({#2})}
\newcommand{\EWT}[2]{\lambda^{\theta}_{#1}({#2})}
\newcommand{\EWDN}[2]{\lambda^{\Dir,\Neu}_{#1}({#2})}
\newcommand{\qD}[1]{\,q^\Dir_{#1}}          
\newcommand{\qN}[1]{\,q^\Neu_{#1}}          

\newcommand{\Ci} [1]{C^\infty ({#1})}       
\newcommand{\Cci}[1]{C_{\mathrm c}^\infty ({#1})} 
\newcommand{\Lsqr}[1]{L_2({#1})}            
\newcommand{\Lsqrloc}[1]{L_{2,\mathrm{loc}}({#1})} 
\newcommand{\Sob}[2][1]{\HS^{#1}({#2})}     
\newcommand{\Sobn}[2][1]{{\overcirc {\mathcal H}{}^{#1}({#2})}}
\newcommand{\norm}[2][{}]{\|{#2}\|_{#1}}    
\newcommand{\normsqr}[2][{}]{\|{#2}\|^2_{#1}} 
\newcommand{\iprod}[3][{}]{\langle{#2},{#3}\rangle_{#1}}    
\newcommand{\bd}  {\partial}                
\newcommand{\orth}{\bot}                    
\newcommand{\map}[3]{{#1}\colon{#2}\longrightarrow{#3}} 
\newcommand{\set}[2]{\{ \, #1 \, | \, #2 \, \} } 

\begin{document}

\title{Eigenvalues in Spectral Gaps of a Perturbed Periodic Manifold}

\author{Olaf Post}
\address{Institut f\"ur Reine und Angewandte Mathematik,
       Rheinisch-Westf\"alische Technische Hochschule Aachen,
       Templergraben 55,
       52062 Aachen,
       Germany}
\email{post@iram.rwth-aachen.de}
\date{05.12.2001}



\begin{abstract}
  We consider a non-compact Riemannian periodic manifold such that the
  corresponding Laplacian has a spectral gap. By continuously perturbing the
  periodic metric locally we can prove the existence of eigenvalues in a gap.
  A lower bound on the number of eigenvalue branches crossing a fixed level is
  established in terms of a discrete eigenvalue problem. Furthermore, we
  discuss examples of perturbations leading to infinitely many eigenvalue
  branches coming from above resp.\ finitely many branches coming from below.
\end{abstract}

\maketitle


\section{Introduction}

A \emph{periodic manifold} is a (non-compact) Riemannian manifold $\Mper$ with
metric $g$ such that a finitely generated Abelian group $\Gamma$ acts properly
discontinuously and isometrically on $\Mper$. Furthermore, we assume that the
quotient $\Mper/\Gamma$ is compact. We call the closure of a fundamental
domain a \emph{period cell}. We suppose that the Laplacian $\laplacian \Mper$
on $\Mper$ has a spectral gap $I=(a,b)$ in the (essential) spectrum, i.e.,
\begin{displaymath}
  I \cap \spec \laplacian \Mper = \emptyset
\end{displaymath}
where $a > \inf \essspec \laplacian \Mper$.  We prove that a local
perturbation of the periodic structure leads to eigenvalues in the gap for
large deformation parameters $\tau$.  More precisely, we perturb the periodic
metric $g$ smoothly, starting from $g=\pert g 0$ and we denote the manifold
$\Mper$ with perturbed metric $\pert g \tau$ by $\Mpertau$. Furthermore, we
assume that $\pert g \tau$ and its first derivatives depend continuously on
$\tau$ and that the perturbation is (essentially) contained in the compact set
$\Pert$.  Here, $\Pert$ denotes a finite number of copies of a period cell $M$
of $\Mper$.  Later on we allow perturbations $\Mpertau$ diffeomorphic to
$\Mper$ (see Section~\ref{sec:examples}) or even some cases when the
perturbation changes the topology of $\Mper$ (see
Section~\ref{sec:non.diffeo}).

Donnelly~\cite{donnelly:92} and Li~\cite{li:92} proved the existence of
eigenvalues \emph{below} the essential spectrum of the Laplacian on the simply
connected, complete, hyperbolic plane by a local perturbation of the manifold.
The methods employed there use separation of variables. Therefore the ideas
cannot be generalized to our case.  Furthermore the Min-max principle applies,
since the eigenvalues lie below the essential spectrum. Note also the
introductory article~\cite{hislop:94} concerning results on the spectral
theory of hyperbolic manifolds.

\sloppy In general, it is much more difficult to prove the existence of
eigenvalues \emph{in} a spectral gap. Such problems have been extensively
studied in the case of Schr\"odinger or divergence type operators with a
spectral gap (see e.g.~\cite{hempel-deift:86}, \cite{alama-deift-hempel:92},
\cite{aadh:94} or \cite{hempel-besch:00}). The school of Birman (see e.g.
\cite{birman:97} or \cite{safronov:01} and references given therein) obtained
a vast number of results on the asymptotics of eigenvalue counting functions
using the Birman-Schwinger principle. This principle does not apply to our
case since the perturbation is not additive.  General results on perturbation
of point spectra can be found in~\cite{kato:66} or~\cite{reed-simon-4}.

{}From a physical point of view the Laplacian on a periodic manifold is the
Hamiltonian of an electron confined to a periodically curved material,
cf.~\cite{froese-herbst:00} or \cite{mitchell:01}.  For example, consider a
flat small strip in $\R^2$ (i.e., a \emph{quantum wire} or \emph{quantum wave
  guide}) which is periodically curved. If the strip is thin enough, there
exist gaps in the spectrum of the Dirichlet Laplacian
(cf.~\cite{exner-seba:89}~and \cite{yoshitomi:98}).  The local perturbation of
the periodic structure leads to bound states (i.e.\ square integrable
eigenfunctions) in a spectral gap.  Physically, the perturbation corresponds
to local impurities coming from contamination, e.g., in a semi-conductor. In
particular, we show that local deformations of a periodically curved quantum
wire produce bound states with energy in a spectral gap of the Hamiltonian of
the non-perturbed quantum wire (see Proposition~\ref{prop:qw.blow}).

\subsection*{Assumptions}
The main assumption on the periodic manifold $\Mper$ resp.\ the period cell
$M$ is the following. Let $\EWD k M$ and $\EWN k M$ denote the eigenvalues
(written in increasing order and repeated according to multiplicity) of the
Laplacian on the period cell $M$ with Dirichlet resp.\ Neumann boundary
condition on $\bd M$.  Note that we always have $\EWN k M \le \EWD k M$ in
virtue of the Min-max principle (see the next section for details). We assume
that the following \emph{gap condition} is fulfilled:
\begin{equation}
  \label{eq:gap}
  \exists \, k \in \N \colon \qquad 
  I_k:=\bigl( \EWD k M, \EWN {k+1} M \bigr) \ne \emptyset,
\end{equation}
i.e., $\EWD k M < \EWN {k+1} M$ for some $k$. From Floquet theory and the
Min-max principle it follows that $I_k$ lies \emph{in} a spectral gap for the
Laplacian on $\Mper$. Note that $I_k$ could be \emph{strictly} contained in a
spectral gap. The gap condition~\eqref{eq:gap} is fulfilled if the period cell
decouples from its translates, i.e., if the junctions between different period
cells are small in some sense. Recall that the gap condition is \emph{not}
satisfied for $\Mper=\R^d$ with the standard metric.  In this case, we always
have $\EWD k M \ge \EWN {k+1} M$.  In Section~\ref{sec:mfd.gaps} we give
examples of periodic manifolds with an arbitrary (finite) number of gaps that
were obtained in \cite{post:01b} and \cite{yoshitomi:98}.

If $\Gamma'$ denotes a subset of the Abelian group $\Gamma$ we set
\begin{displaymath}
  \GpM := \bigcup_{\gamma \in \Gamma'} \gamma M,
\end{displaymath}
in particular, $\Gamma M = \Mper$.  Consider a sequence $(\Gamma^n)$ of
subsets of $\Gamma$ such that $\Gamma^n$ has $n$ elements. We assume that
$\Gamma^n \nearrow \Gamma$ and set $\Appr := \GnM$, $\Rest := (\Gamma
\setminus \Gamma^m) M$ and $\RestAppr := (\Gamma^n \setminus \Gamma^m) M$. As
for $\Mpertau$ we denote by $\Apprtau$ resp.\ $\RestApprtau$ the manifold
$\Appr$ resp.\ $\RestAppr$ with perturbed metric $\pert g \tau$.

The assumptions on the family $(\pert g \tau)$ of perturbed metrics on $\Mper$
are the following:
\begin{align}
  \label{eq:unperturbed}
  \pert g 0 = g   & \qquad \text{on $\Mper$} \\
  \label{eq:met.cont}
  \norm[C^1] {\pert g \tau - \pert g {\tau_0}} \to 0
  & \qquad\text{on $\Mper$ as $\tau \to \tau_0$}\\
  \label{eq:pert.compact}
  \pert g \tau = \pert g 0 & \qquad\text{on $\Rest$, $\tau \ge 0$} 
\intertext{for all
    $\tau_0 \ge 0$ and sufficiently large $m \in \N$. The $C^1$-norm of a
    metric is defined in~\eqref{eq:norm.met}. Furthermore, if $\bd M$ is
    smooth we can also allow non-compact perturbations which are small outside
    $\Pert$, i.e.,}
  \label{eq:met.unif.rest} \tag{1.4'}
  \sup_{\tau \ge 0} \norm[C^1] {\pert g \tau - \pert g 0} \to 0 
  & \qquad\text{on $\Rest$ as $m \to \infty$.}
\end{align}
Note that $\bd M$ is smooth iff $\bd \Rest$ is smooth.

Assumption~\eqref{eq:unperturbed} assures that we start from the periodic
manifold. Assumption~\eqref{eq:met.cont} guarantees the continuous dependence
of the metric and its first derivatives on $\tau$. In particular, eigenvalues
on the approximating problem $\Apprtau$ depend continuously on $\tau$
(cf.~Corollary~\ref{cor:ew.cont}), and the norm resp.\ weak topologies of the
corresponding Hilbert spaces do not depend on $\tau$
(cf.~Corollary~\ref{cor:norm.quad.est}). Note that we only need to control the
metric up to its first derivatives since no higher derivative of the metric
occurs in the coefficients of the Laplacian,
cf.~Equation~\eqref{eq:lapl.local}.

Finally, Assumption~\eqref{eq:pert.compact} resp.~\eqref{eq:met.unif.rest}
says that the support of the perturbation is contained in the compact set
$\Pert$ resp.\ is small outside the compact set $\Pert$. In particular, one
can prove a \emph{decomposition principle} for non-compact perturbations,
i.e., the invariance of the essential spectrum under such perturbations (see
Theorem~\ref{thm:dec.princ}). With other words, the essential spectrum
reflects the geometry at infinity. In our situation, the decomposition
principle assures that the spectral gaps of $\laplacian \Mper$ remain spectral
gaps in the \emph{essential} spectrum of $\laplacian \Mpertau$ for all $\tau
\ge 0$.  Only discrete eigenvalues (possibly accumulating at the band-edges of
$\spec \laplacian \Mper$) can appear in the spectral gaps.

\subsection*{Main result}
To count the number of eigenvalues $\lambda$ (with multiplicity) of the family
$\laplacian {\pert \Mper {\tau'}}$, $0 \le \tau' \le \tau$, we define
\begin{equation}
  \label{eq:def.count.fct}
  \mathcal N(\tau,\lambda) := 
  \sum_{0 \le \tau' \le \tau} 
       \dim \ker (\laplacian {\pert \Mper {\tau'}} - \lambda).
\end{equation}
Note the difference to the ordinary eigenvalue counting function
\begin{displaymath}
  \dimEWD[I] M := 
  \sum_{\lambda \in I} 
       \dim \ker (\laplacianD M - \lambda)
\end{displaymath}
that counts the eigenvalues in the Borel subset $I \subset \R_+$ of the
Dirichlet Laplacian on $M$; similarly, $\dimEWN[I] M$ denotes the
number of eigenvalues in $I$ of the Neumann Laplacian. As an
abbreviation, we set $\dimEWD M := \dimEWD [{[0,\lambda]}] M$.

Our main result is the following:
\begin{theorem}
  \label{thm:eigenvalues}
  Let $\lambda \in I_k$ belong to a spectral gap of $\laplacian \Mper$ then
  \begin{align}
    \label{eq:from.above}
    \mathcal N(\tau,\lambda) &\ge 
    \dimEWD \Perttau - \dimEWD \Pert\\
    \label{eq:from.below}
    \mathcal N(\tau,\lambda) &\ge 
    \dimEWN \Pert - \dimEWN \Perttau
  \end{align}
  for all $\tau \ge 0$ and sufficiently large $m \in \N$.
\end{theorem}
Note that Theorem~\ref{thm:eigenvalues} reduces the eigenvalue problem on the
non-compact manifold $\Mper$ to one on the compact manifold $\Pert$.
Therefore, we can apply the Min-max principle~\eqref{eq:max.min} to assure the
existence of eigenfunctions of the perturbed problem on $\Perttau$.

In a semi-classical picture, the number of bound states for high energy levels
is approximately given by the phase space volume associated with the classical
energy of the quantum system. The classical Hamiltonian on a $d$-dimensional
Riemannian manifold $X$ corresponding to $\laplacian X$ is given by
$h(x,p):=g^*_x(p,p)$ where $g^*$ denotes the metric on the cotangent bundle
$T^*\! X$ and $p \in T^*_x X$. The Hamiltonian $h$ determines a region in the
phase space $T^*\! X$ in which a classical particle with given energy is
allowed to move. The uncertainty principle, however, demands that each bound
state requires a cube of volume $(2\pi)^d$ in phase space, and therefore the
total number of bound states is approximately equal to the phase space volume
divided by $(2\pi)^d$ (see~\cite{reed-simon-4}). Indeed, the Weyl asymptotic
distribution of the (Dirichlet or Neumann) eigenvalues of the Laplacian on $X$
is given by
\begin{equation}
  \label{eq:weyl}
  \dimEW X - 
  \frac {\omega_d}{(2\pi)^d} \lambda^{\frac d 2}  \, \vol(X) =
  O(\lambda^{\frac{d-1}2}), \qquad \lambda \to \infty
\end{equation}
where $\omega_d$ is the volume of the unit ball in $\R^d$
(see~e.g.~\cite{safarov:98}).  Since we expect that the bound states are
localized on the perturbed manifold $\Pert$ we anticipate that
\begin{equation}
\label{eq:phase.diff}
  \frac {\omega_d}{(2\pi)^d} \lambda^{\frac d 2} |\vol(\Pert) - \vol(\Perttau)|
\end{equation}
is a lower bound for the eigenvalue counting
function~\eqref{eq:def.count.fct}.  However, note that
$O(\lambda^{\frac{d-1}2})$ in~\eqref{eq:weyl} depends on the manifold $\Pert$.
Nevertheless in our examples in Sections~\ref{sec:examples}
and~\ref{sec:non.diffeo} we show that an infinite number of eigenvalue
branches crosses the level $\lambda$ from above if the volume of $\Perttau$
increases. In contrast, if we decrease the volume of $\Perttau$, only a finite
number of eigenvalue branches cross the level $\lambda$ from below. Here, by
an \emph{eigenvalue branch} we mean an eigenvalue of (one of) the perturbed
manifold(s) viewed as a function depending on the perturbation parameter
$\tau$.

Under the assumptions given above it is not possible to obtain an upper bound
on the function $\mathcal N(\tau,\lambda)$. Since the dependence on $\tau$ is
not supposed to be analytic in Assumption~\eqref{eq:met.cont}, we could
consider a family of metrics $(\pert g \tau)$ which is constant near $\tau_0$.
If $\lambda$ is an eigenvalue of $\laplacian{\Mpertau[_0]}$, then $\mathcal
N(\tau_0,\lambda)=\infty$.

Nevertheless, the examples in Sections~\ref{sec:examples}
and~\ref{sec:non.diffeo} in which the eigenvalue branches of the problem on
$\Perttau$ are monotonical should be good candidates where an \emph{upper}
bound on the counting function $\mathcal N(\tau,\lambda)$ in terms of
$| \! \dimEW \Perttau - \dimEW \Pert|$ holds (cf.
Remarks~\ref{rem:mono.conf},~\ref{rem:mono.diff} and~\ref{rem:monotonical}).
Under suitable further conditions one should have to assure that the
eigenvalue branches (in the gap) of the \emph{full} problem $\Mpertau$ are
\emph{strictly} monotonical. Note that --- in contrast to the approximating
problem on $\Perttau$ --- the monotonicity of the eigenvalue branches of the
\emph{full} problem $\Mpertau$ cannot be shown by the Min-max principle since
the gap lies above the infimum of the essential spectrum.

Note furthermore that the counting function defined above could be much
greater than the \emph{spectral flow} of the family $(\laplacian \Mtau)$ at
$\lambda$ defined as the difference of the eigenvalue branches crossing the
level $\lambda$ from above and below (cf.~e.g.~\cite{safronov:01}).  Think for
example of an eigenvalue branch which oscillates around $\lambda$.

In~\cite{hempel-deift:86} or~\cite{alama-deift-hempel:92}) a family
$(\laplacian \Mpertau)$ is called \emph{complete} if for each $\lambda \ge 0$
there exists $\tau \ge 0$ such that $\lambda \in \spec \laplacian \Mpertau$.
Note that all our examples given in Sections~\ref{sec:examples}
and~\ref{sec:non.diffeo} are complete in this sense (in
Proposition~\ref{prop:conf.shrink} we have to perturb at least $m=2$ period
cells to ensure that $\mathcal N(\tau, \lambda) \ge 1$ for $\tau$ large
enough).

For the proof of Theorem~\ref{thm:eigenvalues} we cannot directly apply the
Min-max principle since we consider eigenvalues inside a spectral gap.
Therefore, we adopt the following idea given in~\cite{hempel-deift:86},
\cite{alama-deift-hempel:92}, \cite{aadh:94} or~\cite{hempel-besch:00} to our
situation: we show that the eigenfunctions of the full problem on $\Mper$ can
be approximated by eigenfunctions of an approximating problem on $M^n$
(cf.~Theorem~\ref{thm:ef.conv} and Theorem~\ref{thm:count.approx}). Our
situation differs from the case given in the papers mentioned above: in some
sense it is more complicated since we deal with different Hilbert spaces
$\Lsqr \Mpertau$ for each $\tau$. On the other hand, our situation is simpler
since there are no extra eigenvalues in the gap coming from the boundary
conditions on the approximating problem (cf.~Lemma~\ref{lem:common.gap}).

The paper is organized as follows: In Section~\ref{sec:prelim} we provide some
notation and basic results on Laplacians and periodic manifolds. In
Section~\ref{sec:ell.est} we develop elliptic estimates needed for the proof
of Theorem~\ref{thm:eigenvalues} given in Section~\ref{sec:eigenvalues}. In
Section~\ref{sec:mfd.gaps} we recall examples of periodic manifolds with
spectral gaps. Finally, in Sections~\ref{sec:examples}
and~\ref{sec:non.diffeo} we present examples of diffeomorphic and
non-homeomorphic perturbations leading to eigenvalues.

\section{Preliminaries}
\label{sec:prelim}

\subsection*{Laplacian on a manifold}

Throughout this article we study manifolds of dimension $d \ge 2$. For a
Riemannian manifold $M$ (compact or not) without boundary we denote by $\Lsqr
M$ the usual $L_2$-space of square integrable functions on $M$ with respect to
the volume measure on $M$. In a chart, the volume measure has the
density $(\det g)^\frac12$ with respect to the Lebesgue measure, where $\det
g$ is the determinant of the metric tensor $(g_{ij})$ in this chart. The norm
of $\Lsqr M$ will be denoted by $\norm[M] \cdot$. For $u \in \Cci M$, the
space of compactly supported smooth functions, we set
\begin{displaymath}
  \check q_M(u):=\normsqr[M] {\dd u} = \int_{M} |\dd u|^2.
\end{displaymath}
Here the $1$-form $\dd u$ denotes the exterior derivative of $u$ given in
coordinates by $| \dd u |^2= \sum_{i,j} g^{ij} \partial_i u \, \partial_j
\overline u$ where $(g^{ij})$ is the inverse matrix of $(g_{ij})$.

We denote the closure of the non-negative quadratic form $\check q_M$
by $q_M$.  Note that the domain $\dom q_M$ of the closed quadratic
form $q_M$ consists of functions in $L_2(M)$ such that the weak
derivative $\dd u$ is also square integrable (i.e., $q_M(u) <
\infty$).

We define the \emph{Laplacian} $\laplacian M$ (for a manifold without
boundary) as the unique self-adjoint and non-negative operator
associated with the closed quadratic form $q_M$, i.e., operator and
quadratic form are related by
\begin{displaymath}
  q_M(u)=\iprod {\laplacian M u} u
\end{displaymath}
for $u \in \Cci M$ (for details on quadratic forms see e.g.\ 
\cite[Chapter~VI]{kato:66}, \cite{reed-simon-1} or \cite{davies:96}).
Therefore, the Laplacian for smooth functions $u$ is given in a chart by
\begin{equation}
  \label{eq:lapl.local}
  \laplacian M u = 
  -(\det g)^{-\frac12} \sum_{i,j} \partial_i 
  \bigl(
    (\det g)^\frac12 g^{ij} \, \partial_j u
  \bigr).
\end{equation}

If $M$ is a complete manifold with piecewise smooth boundary $\bd M \ne
\emptyset$ we can define the Laplacian with \emph{Dirichlet} resp.\ 
\emph{Neumann boundary condition} in the same way. Here, we start from the
(closure of the) quadratic form $\check q_M$ defined on $\Cci M$, the space of
smooth functions with compact support \emph{away} from the boundary, resp.\ on
the subspace of $\Ci M$ with $\normsqr[M] u$ and $q_M(u)$ finite. Here, $\Ci
M$ denotes the space of smooth functions with derivatives continuous up to the
boundary of $M$. We denote the closure of the quadratic form by $q_M^\Dir$
resp.\ $q_M^\Neu$ and the corresponding self-adjoint operator by $\laplacianD
M$ resp.\ $\laplacianN M$.  If we are only interested in the differential
expression of the Laplacian, we suppress the boundary condition label. If $\bd
M$ can be split into two disjoint closed sets $\bd_1 M$ and $\bd_2 M$ we
define the Laplacian with mixed boundary conditions,. i.e., Dirichlet boundary
condition on $\bd_1 M$ and Neumann boundary condition on $\bd_2 M$ in the
obvious way.

If $M$ is compact the spectrum of $\laplacian M$ (with any boundary condition
if $\bd M \ne \emptyset$) is purely discrete. Note that we always assume that
$\bd M$ is piecewise smooth. We denote the corresponding eigenvalues by $\EW k
M$, $k \in \N$, (resp.\ $\EWD k M$ or $\EWN k M$ in the Dirichlet or Neumann
case) written in increasing order and repeated according to multiplicity. The
corresponding eigenfunctions are $\Ci M$ up to the boundary.  With this
eigenvalue ordering, we can state the \emph{Min-max principle} (for this
version of the Min-max principle see~e.g.~\cite{davies:96}). The $k$-th
eigenvalue of $\laplacianD M$ can be expressed by
\begin{equation}
  \label{eq:max.min}
  \EWD k M = 
  \inf_{L_k} \sup_{u \in L_k, u \ne 0} 
      \frac {\normsqr{\dd u}}{\normsqr u}
\end{equation}
where the infimum is taken over all $k$-dimensional subspaces $L_k$ of $\dom
q_M^\Dir$. Similar results hold for Neumann or other boundary conditions.

Suppose now that $M=M_1 \cup M_2$ such that $M_1$ and $M_2$ have piecewise
smooth boundary and that $M_1 \cap M_2$ has measure $0$. Denote by $\EWN k
{M_1 \dot \cup M_2}$ the eigenvalues of the quadratic form $q^\Neu_{M_1}
\oplus q^\Neu_{M_2}$ and similarly for Dirichlet boundary condition. Then the
Min-max principle implies
\begin{equation}
  \label{eq:dir.neu.brack}
  \EWN k {M_1 \dot \cup M_2} \le 
  \EWN k M \le 
  \EWD k M \le 
  \EWD k {M_1 \dot \cup M_2}
\end{equation}
for all $k \in \N$ since the opposite inclusions hold for the corresponding
quadratic form domains.  This estimate is called \emph{Dirichlet-Neumann
  bracketing}.

\subsection*{Periodic manifolds and Floquet Theory}

Let $\Gamma$ be an Abelian group of infinite order with $r$ generators and
neutral element $1$.  Such groups are isomorphic to $\Z^{r_0} \times
\Z_{p_1}^{r_1} \times \dots \times \Z_{p_a}^{r_a}$ with $r_0>0$ and $r_0 +
\dots + r_a=r$.  Here, $\Z_p$ denotes the Abelian group of order $p$.  A
$d$-dimensional (non-compact) Riemannian manifold $\Mper$ will be called
\emph{periodic} (or \emph{covering manifold}) if $\Gamma$ acts properly
discontinuously, isometrically and cocompactly on $\Mper$. Cocompactness means
that the quotient space $\Mper / \Gamma$ is compact. Note that the quotient
space is a $d$-dimensional compact Riemannian manifold and that the quotient
map $\map \pi \Mper {\Mper/\Gamma}$ is a local isometry. The metric turning
the manifold $\Mper$ into a periodic one will also be called \emph{periodic}.
For details on periodic manifolds see~e.g.~\cite{chavel:93}.

A compact subset $M$ of $\Mper$ is called a \emph{period cell} if $M$ is the
closure of a fundamental domain $D$, i.e., $M=\overline D$, $D$ is open and
connected, $D$ is disjoint from any translate $\gamma D$ for all $\gamma \in
\Gamma$, $\gamma \ne 1$, and the union of all translates $\gamma M$ is equal
to $\Mper$. Furthermore we assume that $M$ has piecewise smooth boundary.

Floquet theory allows us to analyse the spectrum of the Laplacian on $\Mper$
by analysing the spectra of Laplacians with quasi-periodic boundary condition
on a period cell $M$. In order to do this, we define $\theta$-periodic
boundary condition. Let $\theta$ be an element of the dual group $\hat \Gamma
= \Hom(\Gamma,\Torus^1)$ of $\Gamma$, which is isomorphic to a closed subgroup
of the $r$-dimensional torus $\Torus^r = \{ \theta \in \C^r; \, |\theta_i|=1
\text{ for all $i$} \}$. We denote by $q_M^\theta$ the closure of the
quadratic form $\check q_M$ defined on the space of those functions $u \in \Ci
M$ that satisfy
\begin{displaymath}
  u(\gamma x)= \overline{\theta(\gamma)} \, u(x)
\end{displaymath}
for all $x \in \bd M$ and all $\gamma \in \Gamma$ such that $\gamma x \in \bd
M$. The corresponding operator is denoted by $\laplacianT M$. Again,
$\laplacianT M$ has purely discrete spectrum denoted by $\EWT k M$. The
eigenvalues depend continuously on $\theta$
(see~\cite{reed-simon-4} or~\cite{bratteli:99}).  From Floquet theory we
obtain
\begin{displaymath}
  \spec \laplacian \Mper = \bigcup_{\theta \in \hat \Gamma} \spec \laplacianT M
                         = \bigcup_{k \in \N} B_k(\Mper)
\end{displaymath}
where $B_k=B_k(\Mper) = \{ \EWT k M ; \, \theta \in \hat \Gamma\}$ is a
compact interval, called the \emph{$k$-th band} (see e.g. \cite{reed-simon-4},
\cite{donnelly:81}) or \cite{bruening:92}.  Note that we have the following
\emph{Dirichlet-Neumann enclosure}
\begin{equation}
  \label{eq:dir.neu.encl}
  \EWN k M \le \EWT k M \le \EWD k M
\end{equation}
which follows easily from the opposite inclusions of the corresponding
quadratic form domains via the Min-max principle. Therefore, the $k$-th
Dirichlet resp.\ Neumann eigenvalue are an upper resp.\ lower bound for the
$k$-th band. In particular, if the interval $I_k$ defined in~\eqref{eq:gap} is
non-empty, it lies in a spectral gap of $\laplacian \Mper$. In general, we do
not know whether the intervals $I_k$ are empty or not. In many cases, the
Dirichlet-Neumann enclosure is too rough to guarantee the gap
condition~\eqref{eq:gap}.  Nevertheless, in Section~\ref{sec:mfd.gaps} we cite
examples where the gap condition holds.

\section{Elliptic estimates and perturbations of the metric}
\label{sec:ell.est}

First, we construct an atlas of $\Mper$ adapted to the periodic
structure.  This enables us to define a ``natural'' norm on the
Sobolev space on the non-compact periodic manifold $\Mper$. A more
general concept requires only bounds on the Ricci curvature instead of
the periodicity (cf.~e.g.~\cite{hebey:96}). Next, we need some
elliptic estimates. These estimates are used to obtain estimates in which
different Laplacians (defined with respect to different metrics) occur.
 
\subsection*{Periodic atlases and Sobolev spaces} Suppose that $(\widetilde
U_\beta)$, $\beta \in B$, is a finite cover of $\Mper / \Gamma$ and that
$\map{\widetilde \varphi_\beta}{\widetilde U_\beta}{\widetilde V_\beta}$ are
charts with open sets $\widetilde V_\beta \subset \R^d$. We assume that all
transition maps $\widetilde \varphi_{\beta_1} \circ \widetilde
\varphi_{\beta_2}^{-1}$ (when defined) have bounded derivatives up to all
orders.  We can lift this atlas of the quotient to an atlas
$\map{\varphi_\alpha}{U_\alpha}{V_\alpha}$, $\alpha:=(\beta,\gamma) \in A:=B
\times \Gamma$ of the periodic manifold $\Mper$, i.e., $\widetilde
\varphi_\beta \circ \pi = \varphi_\alpha$, $U_{(\beta,\gamma)} = \gamma
U_{(\beta,1)}$ and $\widetilde V_\beta = V_\alpha$. We call such an atlas $A$
\emph{periodic}.

If we consider $\GpM$ instead of the full periodic manifold $\Gamma M=\Mper$
we set $V'_\alpha := \varphi_\alpha(U_\alpha \cap \GpM)$ which is an open set
in $\R_+^d := [0,\infty) \times \R^{d-1}$. Note that if $\Gamma' \ne \Gamma$
then $\GpM$ has non-empty boundary being isometric to (copies of) $\bd M$. We
also call the atlas $A':=B \times \Gamma'$ of $\GpM$ with charts
$\map{\varphi_\alpha}{U_\alpha \cap \GpM}{V'_\alpha}$ \emph{periodic}.

If $\widetilde g_\beta$ denotes the metric of $\Mper / \Gamma$ carried via the
chart $\widetilde \varphi_\beta$ on $\widetilde V_\beta$ and $g_\alpha$ the
periodic metric $g$ of $\Mper$ carried via $\varphi_\alpha$ on
$V_\alpha=\widetilde V_\beta$ then $\widetilde g_\beta=g_\alpha$ for all
$\gamma \in \Gamma$, i.e., the set of different metrics $\set{g_\alpha}{a \in
  A}$ has only $|B| < \infty$ many elements although $A$ is not finite.

In the same way we can lift a partition of unity $(\widetilde \chi_\beta)$
subordinated to the cover $(\widetilde U_\beta)$ to a partition of unity
$(\varphi_\alpha)$ subordinated to $(U_\alpha)$, i.e., we have $\widetilde
\chi_\beta \circ \pi = \chi_\alpha$. We call such a partition of unity
\emph{periodic}. Again, the set of different functions $\set{\chi_\alpha \circ
  \varphi_\alpha^{-1}}{a \in A}$ is finite.

The next definitions are useful for the elliptic estimate in
Theorem~\ref{thm:reg.theory}.
\begin{definition}
  \label{def:unif.ell}
  A metric $h$ on $\GpM$ (not necessarily periodic) will be called
  \emph{uniformly elliptic} (with respect to the atlas $A'$) if there exists a
  constant $c>0$ such that
  \begin{equation}
    \label{eq:unif.ell}
    c^{-1} |v|^2 \le h_\alpha(x)(v,v) \le c |v|^2
  \end{equation}
  for all $v \in \R^d$, all $x \in V'_\alpha$ and all $\alpha \in A'$. We
  will also write $c^{-1} \le h \le c$ for short.
\end{definition}
Next, we define a norm on the space of $C^1$-sections in the bundle of
symmetric bilinear sections on $\GpM$. A Riemannian metric is an element of
this space.
\begin{definition}
  \label{def:norm.met}
  A metric $h$ will be called \emph{$C^1$-bounded} on $\GpM$ (with respect to
  the atlas $A'$) if
  \begin{equation}
    \label{eq:norm.met}
    \norm[C^1] h := 
    \sup_\alpha \norm[C^1] {h_\alpha} := 
    \sup_\alpha \sup_{x \in V'_\alpha} \max_{i,j,k} 
          \{|h_{\alpha,ij}(x)|, |\partial_k h_{\alpha,ij}(x)| \}
  \end{equation}
  is finite.
\end{definition}
We let $\mathcal G(c,c_1,\Gamma')$ be the space of all uniformly elliptic and
$C^1$-bounded metrics $h$ on $\GpM$ such that $c^{-1} \le h \le c$ and
$\norm[C^1] h \le c_1$.

\begin{remark}
  \label{rem:ex.ell.met}
  Note that the atlas can always be chosen in such a way that a periodic
  metric $g$ on $\Mper$ is uniformly elliptic and $C^1$-bounded. If necessary
  we slightly need to make the charts smaller such that $g_\alpha$ has
  derivatives continuous up to the boundary of $V_\alpha$. In the same way,
  any metric $h$ on $\GpM$ is uniformly elliptic if $\Gamma'$ is finite.
\end{remark}
\begin{remark}
  It seems to be unsatisfactory that our definitions depend on the atlas $A'$.
  We also could define the uniform ellipticity in a coordinate free manner by
  comparing a metric $h$ with the periodic metric $g$ globally.  Likewise, we
  could define the $C^1$-boundedness in a global way by defining how to
  derivative a metric. But since we have chosen a periodic atlas one can easily
  see that the local and global definitions are equivalent. For a similar
  concept of $C^1$-convergence of manifolds
  see~e.g.~\cite{hebey:96}.
\end{remark}

For $k \in \N$ and an open set $V' \subset \R^d_+$ let $\Sob[k] {V'}$ be the
usual Sobolev space of order $k$ defined as the closure of $\Ci {V'}$ under
the norm
\begin{displaymath}
  \normsqr[k, V'] u :=
  \sum_{|\kappa| \le k}\normsqr[V']{\partial_\kappa u}.
\end{displaymath}
We define the Sobolev space $\Sob[k] \GpM$ on $\GpM$ as the closure of the
space of all functions $u \in \Ci \GpM$ such that
\begin{equation}
  \label{eq:sob.norm}
  \normsqr[k, \GpM] u := 
  \sum_{\alpha \in A'} \normsqr[k, V'_\alpha] {u_\alpha}
\end{equation}
is finite.  Here, $u_\alpha := (\chi_\alpha u) \circ \varphi_\alpha^{-1}$
denotes the local representation of $\chi_\alpha u$ in the chart
$\varphi_\alpha$.  We denote by $\Sobn[k] \GpM$ the completion of $\Cci \GpM$,
i.e., the space of smooth functions with compact support \emph{away} from $\bd
\GpM$, under the norm defined in \eqref{eq:sob.norm}.

Since the periodic metric $g$ is uniformly elliptic, we conclude
\begin{equation}
  \label{eq:norm.equiv}
  c^{-d/4} \norm[\GpM] u \le
  \norm[0, \GpM] u =
  \sum_\alpha \norm[V_\alpha] {u_\alpha} \le
  c^{+d/4}|B|(2r+1) \, \norm[\GpM] u,
\end{equation}
i.e., $\norm[\GpM] \cdot$ and $\norm[0, \GpM] \cdot$ are equivalent
norms.  Remember that $r$ denotes the number of generators of $\Gamma$. Note
that a chart $U_\alpha$ could intersect with at most $|B|(2r+1)$ other charts.
In the same way we can show that $(\normsqr u + \normsqr{\dd u})^\frac12$ and
$\norm[1, \GpM] u$ are equivalent norms on $\Sob \GpM$. In particular, we
have $\Sobn \GpM = \dom \qD \GpM$ and $\Sob {\GpM} = \dom \qN {\GpM}$.

\subsection*{Elliptic estimates}

The main result of this subsection is the following theorem:
\begin{theorem}
  \label{thm:reg.theory}
  For every $c > 0$, $c_1 > 0$ and $d > 0$ there exists a constant $c_2$ such
  that
  \begin{equation}
    \label{eq:reg.theory}
    \norm[2, U] u \le c_2 \, (\norm[\GpM] u + \norm[\GpM]{\Delta'u})
  \end{equation}
  for all $u \in \Sob[2] U$, all metrics $g' \in \mathcal G(c,c_1,\Gamma')$,
  all open sets $U \subset \GpM$ such that $\dist(U,\bd \GpM) \ge d$ and all
  $\Gamma' \subset \Gamma$. Here, $\Delta'$ denotes the Laplacian on $\GpM$
  with respect to the metric $g'$.
 
  If $\bd M$ is smooth Estimate~\eqref{eq:reg.theory} is valid for $U=\GpM$,
  all $u \in \Sob[2] \GpM \cap \Sobn \GpM$, all metrics $g' \in \mathcal
  G(c,c_1,\Gamma')$ and all $\Gamma' \subset \Gamma$.
\end{theorem}
\begin{proof}
  The proof of the statement in the chart $V'_\alpha$ is standard (see e.g.\ 
  \cite{gilbarg-trudinger:77}). The step from the local to the global estimate
  is possible since the constant in the local estimate has a global bound.
  Here, we need Estimate~\eqref{eq:norm.equiv} and therefore the special
  structure of the atlas and the partition of unity $(\chi_\alpha)$.
  Furthermore, we have to estimate the first order operators
  $[\Delta',\chi_\alpha]$ in terms of the right hand side
  of~\eqref{eq:reg.theory}. This can be done applying the Gau{\ss}-Green
  formula. Note that $u$ vanishes on the boundary.
\end{proof}

The proof of the next lemma is straightforward. Again, we need
\eqref{eq:norm.equiv} for the step from the local to the global estimate.

\begin{lemma}
  \label{lem:lapl.est.sob}
  There exists a constant $c_3>0$ such that
  \begin{align*} 
    \norm[\Pert] {\laplacian \Appr u} & \le c_3 \, \norm[2, \Pert] u
    \intertext{for all $u \in \Sob[2] \Appr$ and all $m, n$ with $n > m$.
      Suppose further that $\pert g \tau$, $\tau \ge 0$, are metrics on
      $\Mper$ such that $\norm[C^1] {\pert g 0 - \pert g \tau} \to 0$ on
      $\RestAppr$ as $m \to \infty$ ($n > m$) uniformly in $\tau \ge 0$.  Then
      there exists a sequence $\delta'_m \to 0$ such that}
    \norm[\RestAppr]{(\laplacian {\pert \Appr 0} - \laplacian \Apprtau)u} &\le
    \delta'_m \, \norm[2, \RestAppr] u
  \end{align*}
  for all $u \in \Sob[2] \RestAppr$, all $\tau \ge 0$ and all $m,n$ with $n >
  m$.
\end{lemma}

Let $m_+$ be the smallest integer such that $\Pert \subset \overcirc
M{}^{m_+}$. Now we prove the estimates needed in the next section:
\begin{lemma}
  \label{lem:lapl.est}
  Let the family $(\pert g \tau)$ satisfy Conditions~\eqref{eq:unperturbed}
  and~\eqref{eq:met.cont}. Suppose further that $\tau_n \to \tau_0 \ge 0$.
  Then there exists a constant $c_4 > 0$ such that
  \begin{align}
    \label{eq:lapl.est}
    \norm[\Pert]{\laplacian \Appr u} & \le c_4 \, (\norm[\Pertplus] u +
    \norm[\Pertplus]{\laplacian {\Apprtau[_n]} u})
    \intertext{for all $u \in
      \Sob[2] \Appr$ and all $m, n$ with $n > m_+$. Furthermore if in addition
      $\bd M$ is smooth and if Condition~\eqref{eq:met.unif.rest} is satisfied
      then there exists a sequence $\delta_m \to 0$ such that}
    \label{eq:lapl.diff}
    \norm[\RestAppr]{(\laplacian \Appr - \laplacian {\Apprtau[_n]})u} & \le
    \delta_m\, (\norm[\RestAppr] u + \norm[\RestAppr]{\laplacian
      {\Apprtau[_n]} u}) 
  \end{align}
  for all $u \in  \Sob[2] \RestAppr \cap \Sobn \RestAppr$ and all
  $m,n$ with $n > m$ and $m$ large enough.
\end{lemma}

\begin{proof}
  For the proof of \eqref{eq:lapl.est} we combine Lemma~\ref{lem:lapl.est.sob}
  with Theorem~\ref{thm:reg.theory}. Here we need a cut-off function $\chi \in
  \Cci \Pertplus$ such that $\chi \restriction \Pert=1$. Therefore
  $\dist(\supp \chi, \bd \Pert) = d >0$, i.e., we do not need the assumption
  that $\bd M$ is smooth in this case. By Remark~\ref{rem:ex.ell.met}, there
  exist $c,c_1 > 0$ such that $\pert g {\tau_0} \in \mathcal
  G(c,c_1,\Gamma^{m_+})$. Furthermore, Condition~\eqref{eq:met.cont} assures
  that there exist constants $c', c_1' > 0$ such that $\pert g {\tau_n} \in
  \mathcal G(c',c_1',\Gamma^{m_+})$ for all $n \in \N$. Note that we still
  have to estimate the first order operator $[\chi, \laplacian
  {\Apprtau[_n]}]$ in term of the right hand side of~\eqref{eq:lapl.est} as in
  the proof of Theorem~\ref{thm:reg.theory}.  Here we also need the fact that
  $\pert g {\tau_n} \in \mathcal G(c',c_1',\Gamma^{m_+})$.
  
  The proof of \eqref{eq:lapl.diff} is similar. Note that $\pert g {\tau_n}$,
  $n \in \N$, are close to the periodic metric $g=g_0$ by
  Condition~\eqref{eq:met.unif.rest} on $\RestAppr$ provided $m$ is large
  enough. Since the periodic metric $g$ is uniformly elliptic, there exist
  constants $c, c_1 > 0$ such that $\pert g {\tau_n} \in \mathcal
  G(c,c_1,\Gamma^n \setminus \Gamma^m)$ for all $n > m$.
\end{proof}

\subsection*{Perturbation of the metric}
Here, we state some results on how to deal with the different Hilbert
spaces $\Lsqr \Mpertau$ resp.\ $\Sob \Mpertau$ depending on $\tau$. In
particular, we show that the norm and weak topology on $\Lsqr
\Mpertau$ resp.\ $\Sob \Mpertau$ are equivalent for all $\tau \ge
0$; the same is true on the submanifolds $\GpM$
(cf.~Corollary~\ref{cor:norm.quad.est}). Furthermore, we need the
continuity of the eigenvalues with respect to $\tau$,
see~Corollary~\ref{cor:ew.cont} and Lemma~\ref{lem:ew.rest}.

\begin{lemma}
  \label{lem:norm.quad.approx}
  Suppose that the family $(g_\tau)$ satisfies Condition~\eqref{eq:met.cont}.
  Suppose furthermore that $\tau_n \to \tau \ge 0$. Then there exists a
  sequence $\eta_n \to 0$ such that
  \begin{align}
    \label{eq:norm.approx}
    \bigl| \normsqr[\Mpertau] u - \normsqr[{\Mpertau[_n]}] u \bigr| & \le
    \eta_n \, \normsqr[\Mpertau] u\\
    \label{eq:quad.approx}
    \bigl| \normsqr[\Mpertau]{\dd u} - \normsqr[{\Mpertau[_n]}] {\dd u} 
    \bigr| & \le
    \eta_n \, \normsqr[\Mpertau]{\dd u}
  \end{align}
  for all $u \in \Lsqr \Mper$ resp.\ $u \in \Sob \Mper$.
\end{lemma}
\begin{proof}
  We only give the idea of how to prove~\eqref{eq:quad.approx}. The other
  inequality can be proven similarly. We have
  \begin{multline*}
    \int_{\Mpertau} |\dd u|^2 - \int_{\Mpertau[_n]} |\dd u|^2 = \\
    = \sum_\alpha \int_{V_\alpha} \chi_\alpha 
      \Bigl( 
        {\Bigl( E -
            \Bigl( \frac {\det G_{\alpha,n}} 
                         {\det G_\alpha} 
            \Bigr)^\frac12
              G_\alpha^{\frac12}  G_{\alpha,n}^{-1} 
              G_\alpha^{\frac12}
         \Bigr) G_\alpha^{-\frac12} \nabla u} 
        \, \cdot \, {G_\alpha^{-\frac12} \nabla u} 
      \Bigr) (\det G_\alpha)^\frac12.
  \end{multline*}
  Here, $G_\alpha$ resp.\ $G_{\alpha,n}$ denotes the matrix corresponding to
  $\pert g \tau$ resp.\ $\pert g {\tau_n}$ in the chart indexed by $\alpha$.
  Furthermore, $E$ denotes the unit $(d \times d)$-matrix. Clearly,
  Estimate~\eqref{eq:quad.approx} follows from Condition~\eqref{eq:met.cont}.
\end{proof}

The next corollary follows easily. 
\begin{corollary}
  \label{cor:norm.quad.est}
  For all $\tau_0 \ge 0$ there exists a constants $c_5=c_5(\tau_0) \ge 1$ such
  that
  \begin{gather}
    \label{eq:norm.est}
    c_5^{-1} \, \normsqr[\pert \Mper 0] u  \le
                \normsqr[\Mpertau] u  \le
    c_5 \,      \normsqr[\pert \Mper 0] u \\
    \label{eq:quad.est}
    c_5^{-1} \, \normsqr[\pert \Mper 0]   {\dd u}  \le
                \normsqr[\Mpertau]{\dd u}  \le
    c_5      \, \normsqr[\pert \Mper 0]   {\dd u}
  \end{gather}
  for all $0 \le \tau \le \tau_0$.
\end{corollary}

{}From Lemma~\ref{lem:norm.quad.approx} and the Min-max principle we conclude
the following corollary.
\begin{corollary}
  \label{cor:ew.cont} The eigenvalue branch $\tau \mapsto \EWD k
  \Apprtau$ is continuous for $\tau \ge 0$. In particular, for every
  $\tau_0 \ge 0$ there exist a modul of continuity for the eigenvalue
  branch on $[0,\tau_0]$, i.e.\ a non-negative monotonical increasing
  function $\eta$ with $\eta(\delta) \to 0$ as $\delta \to 0$ such
  that \begin{displaymath} |\EWD k \Apprtau - \EWD k {\Apprtau[']}|
  \le \eta(|\tau - \tau'|) \end{displaymath} for all $\tau,\tau' \in
  [0,\tau_0]$.
\end{corollary}

In the same way we can show that the boundary conditions on $\RestApprtau$
have nearly no influence:
\begin{lemma}
  \label{lem:ew.rest}
  Suppose that the family $(g_\tau)$ of metrics satisfies
  Condition~\eqref{eq:met.unif.rest}. Then there exists a sequence $\eta_m \to
  0$ as $m \to \infty$ such that
  \begin{displaymath}
    |\EWD k \RestApprtau - \EWD k \RestAppr| \le \eta_m
  \end{displaymath}
  for all $\tau \ge 0$ and $n > m$. The same estimate is true for Neumann or
  mixed boundary conditions on $\RestAppr$.
\end{lemma}

\section{Eigenvalues in Gaps}
\label{sec:eigenvalues}

In this section we prove our main result Theorem~\ref{thm:eigenvalues}. We
approximate eigenfunctions of the perturbed Laplacian on the full manifold
$\Mper$ by eigenfunctions of the perturbed Dirichlet-Laplacian on the
approximating manifold $\Appr$. This idea has already been used in
\cite{hempel-deift:86}, \cite{alama-deift-hempel:92} or \cite{aadh:94}. Our
situation is in some sense simpler: the approximating and the limit problem
have the same spectral gap.  Otherwise our situation is more difficult since
not only the Laplacian but also the Hilbert space depend on the perturbation
parameter. Some of the technical details are already proven in
Section~\ref{sec:ell.est}.

In Theorem~\ref{thm:ef.conv} we show the convergence of the
approximating eigenfunctions. Furthermore, in Theorem~\ref{thm:count.approx}
we give a lower bound on the eigenvalue counting function defined
in~\eqref{eq:def.count.fct} following arguments of~\cite{hempel-besch:00}.

Firstly, we use the decomposition principle
(see~\cite{donnelly:79}) to prove that the essential spectrum remains
invariant under the perturbation.  Therefore, $\laplacian \Mper$ and
$\laplacian {\Mpertau}$ have the same spectral gap. In a spectral
gap of the unperturbed Laplacian, the perturbed Laplacian can only
have discrete eigenvalues (possibly accumulating at the band edges).
It is essential here that the perturbation is (almost) localized on a
compact set.
\begin{theorem}
  \label{thm:dec.princ}
  We have $\essspec \laplacian \Mper = \essspec \laplacian \Mpertau$
  for all $\tau \ge 0$.
\end{theorem}

\begin{proof}
  We only show the inclusion ``$\subset$'' since the opposite inclusion can be
  proven in the same way.
  
  Let $(u_n)$ be a singular sequence for $\lambda \in \essspec \laplacian
  \Mper$, i.e., $u_n \to 0$ weakly in $\Lsqr \Mper$, $\norm[\Mper]{u_n} = 1$
  and $\norm[\Mper]{(\laplacian \Mper - \lambda)u_n} \to 0$ as $n \to \infty$.
  By the decomposition principle (cf.~\cite{donnelly:79} or \cite{post:00}) we
  can assume that $u_n$ has support away from $\Pert$. Furthermore, by
  Corollary~\ref{cor:norm.quad.est} there exists $c_5>0$ such that
  $\norm[\Mpertau]{u_n} \ge c_5^{-1}$ for all $n$.
  
  Now we want to show that $v_n := u_n / \norm[\Mpertau] {u_n}$ is a singular
  sequence for $\lambda$ and $\laplacian \Mpertau$. Clearly, $v_n \to 0$
  weakly in $\Lsqr \Mpertau$. Since $u_n \restriction \Pert = 0$ we have
  \begin{align*}
    \norm[\Mpertau] {(\laplacian \Mpertau - \lambda) v_n} 
              & \le
    c_5 \, \bigl( 
      \norm[\Rest] {(\laplacian \Mpertau - \laplacian \Mper) u_n} +  
      \norm[\Rest] {(\laplacian \Mper    - \lambda) u_n}  
    \bigr) \\ & \le
    \delta_m \, c_5 \, 
      (\norm[\Rest] {u_n} + \norm[\Rest] {\laplacian \Mper u_n}) + 
      c_5 \norm[\Rest] {(\laplacian \Mper    - \lambda) u_n}  
  \end{align*}
  by an estimate similar to~\eqref{eq:lapl.diff}. Here, $\delta_m \to 0$ as $m
  \to \infty$.  Note that we do not need the smoothness of $\bd M$ here even
  if the perturbation lives on the whole manifold since $\supp u_n$ is
  compactly contained in $\Rest$. Therefore the first term converges to $0$.
  The last term converges to $0$ since $(u_n)$ is a singular sequence for
  $\laplacian \Mper$.
\end{proof}

Next, we conclude from the gap condition~\eqref{eq:gap} and the
Dirichlet-Neumann enclosure~\eqref{eq:dir.neu.encl} that no eigenvalue of the
approximating problem lies in the gap. The boundary of $M$ resp.\ $\GpM$ is so
small such that boundary conditions almost have no influence on the
eigenvalues.
\begin{lemma}
  \label{lem:common.gap}
  Suppose that $\Gamma'$ has $n$ elements.  If $\EWD k M < \EWN {k+1} M$ then
  the interval $I_k=(\EWD k M, \EWN {k+1} M)$ is a common spectral gap, i.e.,
  \begin{equation}
    \label{eq:common.gap}
    I_k \cap \spec \laplacian \Mper = \emptyset \quad \text{and} \quad
    I_k \cap \spec \laplacian \GpM   = \emptyset.
  \end{equation}
  Here, $\laplacian \GpM$ denotes the Laplacian on $\GpM$ with Neumann,
  Dirichlet or mixed boundary condition on $\bd \GpM$. Furthermore, if
  $\lambda \in I_k$ then $\dimEW \GpM = k n$ (also for any boundary
  condition).
\end{lemma}

\begin{proof}
  Denote by $\Sigma^n \! M$ the disjoint union of $n$ copies of $M$. Then
  $\EWD j {\Sigma^n \! M} = \EWD k M$ and $\EWN j {\Sigma^n \!  M} = \EWN k M$
  for all $j=(k-1)n+1, \dots, k n$ since every eigenvalue on the disjoint
  union has multiplicity $p n$ if the corresponding eigenvalue on $M$ has
  multiplicity $p$.  By the Dirichlet-Neumann
  bracketing~\eqref{eq:dir.neu.brack} we obtain
  \begin{displaymath}
    \EWN k M =
    \EWN j {\Sigma^n \! M} \le
    \EW  j \GpM \le
    \EWD j {\Sigma^n \! M} =
    \EWD k M.
  \end{displaymath}
  Note that this estimate is true for any boundary condition on $\GpM$.
  Together with the Dirichlet-Neumann enclosure~\eqref{eq:dir.neu.encl} we see
  that $\EWN k M$ resp.\ $\EWD k M$ are lower resp.\ upper bounds for both $\EW
  j \GpM$ and $\EWT k M$. If $\lambda \in I_k$ then $\laplacian \GpM$ has
  exactly $k n$ eigenvalues below $\lambda$.
\end{proof}

Now we prove that approximating eigenfunctions converge to eigenfunctions of
the full problem and that multiplicity is conserved as $n \to \infty$, i.e.,
as the number of copies of period cells increases. Here and later on we
suppress the Dirichlet label at the approximating problem, e.g.\ we write
$\laplacian \Appr$ instead of $\laplacianD \Appr$ until stated otherwise.

The idea of the next theorem is from~\cite{hempel-deift:86},
\cite{alama-deift-hempel:92} or \cite{aadh:94}:
\begin{theorem}
  \label{thm:ef.conv}
  Suppose that the gap condition~\eqref{eq:gap} is satisfied, i.e., $I_k$ is a
  spectral gap of the periodic Laplacian $\laplacian \Mper$ for some $k \in
  \N$. Suppose further that the family $(\pert g \tau)$ satisfies
  Conditions~\eqref{eq:unperturbed},~\eqref{eq:met.cont}
  and~\eqref{eq:pert.compact} or, if $\bd M$ is
  smooth,~\eqref{eq:met.unif.rest}. Let $\tau_n \to \tau$, $\lambda_{n,i} \to
  \lambda \in I_k$ for all $i=1, \dotsc, j$.  Furthermore, suppose that
  $\varphi_{n,1}, \dots, \varphi_{n,j} \in \Lsqr {\Apprtau[_n]}$ are
  orthonormal for all $n \in \N$ and that
  \begin{equation}
    \label{eq:ew.approx}
    \laplacian {\Apprtau[_n]} \varphi_{n,i} = \lambda_{n,i} \, \varphi_{n,i}. 
  \end{equation}
  Then for every $i$ there exists a subsequence of $(\varphi_{n,i})_n$ also
  denoted by $(\varphi_{n,i})_n$ converging weakly to $\varphi_{0,i} \in \dom
  \laplacian \Mpertau$ in $\Sob \Mpertau$ and strongly in $\Lsqrloc \Mpertau$.
  Furthermore,
  \begin{equation}
    \label{eq:ew.limit}
    \laplacian \Mpertau \varphi_{0,i} = \lambda \, \varphi_{0,i} 
  \end{equation}
  and $\varphi_{0,1}, \dots, \chi_{0,j}$ are linearly independent.
\end{theorem}

\begin{proof}
  Let $i \in \{1, \dots j\}$. From Corollary~\ref{cor:norm.quad.est} it
  follows that $(\varphi_{n,i})_n$ is bounded in $\Sob \Mper$.  Therefore we
  can extract a subsequence also denoted by $(\varphi_{n,i})_n$ converging
  weakly in $\Sob \Mper$ to an element $\varphi_{0,i}$.  This subsequence also
  converges strongly in $\Lsqrloc \Mper$ by the Rellich-Kondrachov compactness
  Theorem.  In virtue of Lemma~\ref{lem:norm.quad.approx} and
  Corollary~\ref{cor:norm.quad.est} it is a straightforward calculation to
  show that $\varphi_{0,i}$ is in the domain of $\laplacian \Mpertau$ and that
  the eigenvalue equation \eqref{eq:ew.limit} is satisfied.
  
  The main difficulty is to prove that $\varphi_{0,i}$ are linearly
  independent. Suppose that there exist $\alpha_1, \dots, \alpha_k$, not all
  equal to $0$,
  such that $u_n := \sum_i \alpha_i \, \varphi_{n,i}$ converges to $\sum_i
  \alpha_i \, \varphi_{0,i} =0$ in $\Lsqrloc \Mpertau$. By
  Lemma~\ref{lem:common.gap} the interval $I_k$ is a spectral gap for all
  operators $\laplacian \Appr$, $n \in \N$. In virtue of the spectral theorem
  and Estimate~\eqref{eq:norm.est} we have
  \begin{equation}
    \label{eq:not.zero}
          \normsqr[\Appr]{(\laplacian \Appr - \lambda) u_n} \ge
    c_6^2 \normsqr[\Appr]{u_n} \ge 
    \frac{c_6^2}{c_5} \sum_i |\alpha_i|^2 > 0
  \end{equation}
  for all $n \in \N$ where $c_6 := \dist(\lambda, \R \setminus I_k)$. On the
  other side we have
  \begin{displaymath}
    \norm[\Appr] {(\laplacian \Appr - \lambda) u_n} \le
    \norm[\Appr] 
        {\sum_i (\laplacian \Appr - \lambda_{n,i}) \alpha_i \, \varphi_{n,i}} +
         \sum_i |\alpha_i| \cdot |\lambda_{n,i}-\lambda| \cdot 
         \norm[\Appr] {\varphi_{n,i}}.
  \end{displaymath}
  The last sum converges to $0$ since $\norm[\Appr] {\varphi_{n,i}} \le c_5$.
  The first norm can be split into an integral over $\Pert$ and $\RestAppr$.
  
  The integral over $\Pert$ can be estimated by
  \begin{displaymath}
    \norm[\Pert] {\laplacian \Appr u_n} + 
    \sum_i \lambda_{n,i} \norm[\Pert] {u_n}. 
  \end{displaymath}
  Furthermore, Inequality~\eqref{eq:lapl.est} yields the estimate
  \begin{multline*}
    \norm[\Pert] {\laplacian \Appr u_n} \le 
    c_4 \, c_5 \, \bigl( \, \norm[\Pertplustau] {u_n} + 
                \norm[\Pertplustau] {\laplacian {\Apprtau[_n]} u_n}
      \bigr) \\ \le
    c_4 \, c_5  \, 
    \bigl( (1 + \sum_i \lambda_{n,i}) 
    \bigr)  \norm[\Pertplustau] {u_n}
  \end{multline*}
  for all $n > m_+$. In the last estimate we have used
  Equation~\eqref{eq:ew.approx}. Note that the last term converges to $0$
  since $u_n \to 0$ in $\Lsqrloc \Mper$. 
  
  If the perturbation is contained in $\Pert$ we have a contradiction
  to~\eqref{eq:not.zero}. Otherwise we still have to estimate the integral
  over $\RestAppr$. Here, we need the smoothness assumption on $\bd M$ to be
  able to use~\eqref{eq:lapl.diff}. Together with~\eqref{eq:ew.approx} we
  obtain
  \begin{multline*}
    \norm[\RestAppr]{(\laplacian \Appr - \laplacian{\Apprtau[_n]})u_n} \le
    \delta_m \, 
    \bigr( \norm[\Appr] {u_n} +
           \norm[\Appr] {\laplacian{\Apprtau[_n]} u_n}
    \bigl)
    \\ \le
    \delta_m \, c_5 \,
    \bigr( \sum_i |\alpha_i|^2 \bigl)^\frac12 (1+\max_i \lambda_{n,i})
  \end{multline*}
  tending to $0$ as $m \to \infty$. Again, we have a contradiction
  to~\eqref{eq:not.zero}.
\end{proof}

The next result allows us to estimate the eigenvalue counting functions for
different parameters $\tau$ and fixed $\lambda$ by the eigenvalue counting
function for a fixed $\tau$ and an interval containing $\lambda$ (see
\cite{hempel-besch:00}). We need this lemma since we do not know whether
eigenfunctions corresponding to different parameters $\tau$ are orthogonal. We
do not even know that they are different! But since we want to show that the
multiplicity of eigenvalues is conserved as $n \to \infty$ we need the linear
independence of the approximating eigenfunctions (see the proof of
Theorem~\ref{thm:count.approx}).
\begin{lemma}
  \label{lem:eigen.vert}
  For all $\tau_0 \ge 0$ there exists a monotonical increasing function
  $\eta(\delta) \to 0$ as $\delta \to 0$ such that
  \begin{displaymath}
    |\dimEWD {\Apprtau[+\delta]}  - 
     \dimEWD {\Apprtau[-\delta]}| \le
     \dimEWD[{[\lambda - \eta(\delta),\lambda + \eta(\delta)]}]
             \Apprtau
  \end{displaymath}
  for all $\lambda \ge 0$, $\delta >0$ and all $\tau + \delta \le \tau_0$.
\end{lemma}
\begin{proof}
  Since every eigenvalue branch $\EWD j {\pert \Appr {(\cdot)}}$ is continuous
  by Corollary~\ref{cor:ew.cont}, there exists a parameter $\tau' \in
  [\tau-\delta,\tau+\delta]$ such that $\lambda=\EWD j {\Apprtau[']}$ by the
  intermediate value theorem. Corollary~\ref{cor:ew.cont} yields
  \begin{displaymath}
    \bigl| \lambda               - \EWD j \Apprtau \bigr| = 
    \bigl| \EWD j {\Apprtau[']}  - \EWD j \Apprtau \bigr| \le
    \eta(|\tau-\tau'|) \le
    \eta(\delta),
  \end{displaymath}
    i.e., $\laplacian \Apprtau$ has an eigenvalue in
  $[\lambda - \eta(\delta),\lambda + \eta(\delta)]$.
\end{proof}

Now we give a lower bound on the eigenvalue counting function
following~\cite{hempel-besch:00}:
\begin{theorem}
  \label{thm:count.approx}
  If $\lambda \in I_k$ lies in a spectral gap then
  \begin{equation}
    \label{eq:count.approx}
    \mathcal N(\tau,\lambda) \ge
    \limsup_{n \to \infty} 
       \bigl|\dimEWD \Apprtau - \dimEWD \Appr \bigr|.
  \end{equation}
  for all $\tau \ge 0$.
\end{theorem}

\begin{proof}
  Denote by
  \begin{displaymath}
    T_n := T_n(\lambda) := \set{\tau' \in [0,\tau_0]} 
                           {\lambda \in \spec \laplacian \Apprtau}, 
  \end{displaymath}
  the set of parameters $\tau'$ that produce an eigenvalue $\lambda$. Let
  $T_\infty$ be the set of limit points, i.e., $\hat \tau \in T_\infty$ if and
  only if $\hat \tau \in [0,\tau_0]$ and if there exist sequences $(n_m)_m
  \subset \N$ and $\tau'_m \in T_{n_m}$ such that $\tau'_m \to \hat \tau$.
  
  We have to distinguish two cases. If the cardinality of $T_\infty$ is
  greater or equal to $N_\lambda$, the right hand side of
  \eqref{eq:count.approx}, then we apply Theorem~\ref{thm:ef.conv} with
  fixed eigenvalue $\lambda=\lambda_{n,1}$ and with multiplicity $j=1$
  for each limit point $\hat \tau \in T_\infty$.  As a consequence, there are
  at least $\card T_\infty$ parameters $\hat \tau$ such that $\lambda$ is an
  eigenvalue of $\Delta_{M(\hat\tau)}$. This proves~\eqref{eq:count.approx}.
  
  If $\card T_\infty < N_\lambda$ then $T_\infty$ consists of a finite number
  of points $\hat \tau_1,\dots,\hat \tau_q$, and $T_n \to \{\hat
  \tau_1,\dots,\hat \tau_q\}$.  Furthermore, there exists a sequence $\delta_n
  \to 0$ such that
  \begin{displaymath}
    T_n \subset 
     \bigcup_{p=1}^q \, (\hat \tau_p-\delta_n,\hat \tau_p+\delta_n) \,
       =: \hat T_n
  \end{displaymath}
  for all $n \in \N$. If $n$ is large enough, all these intervals are mutually
  disjoint. As a consequence, $d_{\lambda,n}(\hat \tau):= \dimEWD {\pert \Appr
    {\hat \tau}}$ is constant on each component of $[0,\tau] \setminus \hat
  T_n$. Therefore
  \begin{displaymath}
    \bigl| d_{\lambda,n}(\tau) - d_{\lambda,n}(0) \bigr| \le
    \sum_{p=1}^q 
      \bigl| d_{\lambda,n}(\hat \tau_p-\delta_n) - 
       d_{\lambda,n}(\hat \tau_p+\delta_n) \bigr|. 
  \end{displaymath}
  By Lemma~\ref{lem:eigen.vert} there exist $\eta(\delta) \to 0$ as
  $\delta \to 0$ such that
  \begin{displaymath}
    \bigl| d_{\lambda,n}(\tau) - d_{\lambda,n}(0) \bigr| \le
    \sum_{p=1}^q j_{p,n} 
    \qquad \text{where} \qquad
    j_{p,n} := 
    \dimEWD[{[\lambda - \eta(\delta_n),\lambda + \eta(\delta_n)]}]
          {\pert \Appr {\hat \tau_p}}.
  \end{displaymath}
  By passing to a subsequence we can assume that $|d_{\lambda,n}(\tau) -
  d_{\lambda,n}(0) | \to N_\lambda$. We have even equality if $n$ is large
  enough since the sequence consists of integers. Now we select another
  subsequence $(n_m)_m$ such that $j_p := j_{p,n_m} \in \N$ is independent of
  $m$. This is possible by choosing a convergent subsequence.  Therefore, for
  each $m \in \N$, we have $j_p$ orthonormal eigenfunctions
  $\varphi_{m,1},\dots,\varphi_{m,j_p}$ of the Laplacian on $\pert \Apprm {\hat
    \tau_p}$ with eigenvalues $\lambda_{m,i} \to \lambda$ as $m \to \infty$.
  Hence we can apply Theorem~\ref{thm:ef.conv} with fixed parameter $\hat
  \tau_p$ and conclude that the limit problem $\laplacian \Mpertau$ has
  $\lambda$ as eigenvalue of multiplicity $j_p$. This proves the theorem since
  $\lambda$ hat multiplicity at least $\sum_{p=1}^q j_p \ge N_\lambda$.
\end{proof}

Finally we want a lower bound on the eigenvalue counting function in
terms of $\Pert$ and \emph{not} in terms of $\Appr$
\begin{lemma}
  \label{lem:approx.probl}
  For $m$ sufficiently large we have
  \begin{gather}
    \label{eq:eigen.count1}
    \dimEWD \Apprtau -
    \dimEWD \Appr \ge 
    \dimEWD \Perttau -
    \dimEWD \Pert\\
    \label{eq:eigen.count2}
    \dimEWD \Appr -  
    \dimEWD \Apprtau \ge
    \dimEWN \Pert -
    \dimEWN \Perttau
   \end{gather}
  for all $\tau \ge 0$ and $n > m$.
\end{lemma}

\begin{proof}
  From Lemma~\ref{lem:common.gap} we know that $\dimEW \Appr= k
  n$ and $\dimEW \RestAppr= k (n-m)$ if $\lambda \in I_k$
  independently of the boundary conditions. Furthermore, from
  Lemma~\ref{lem:ew.rest} we obtain 
  \begin{displaymath}
    \dimEWD \RestApprtau = \dimEWD \RestAppr, \qquad
    n > m, \tau \ge 0
  \end{displaymath}
  if $m$ is large enough.  Furthermore, the Dirichlet-Neumann
  bracketing~\eqref{eq:dir.neu.brack} yields
  \begin{multline*}
    \dimEWD \Apprtau - \dimEWD \Appr \\ \ge
    \dimEWD \Perttau + \dimEWD \RestApprtau -
    \bigl(
      \dimEWD \Pert  + \dimEWND \RestAppr
    \bigr) \\ =
    \dimEWD \Perttau - k m.
  \end{multline*}
  where $\dimEWND \RestAppr$ denotes the eigenvalue counting function of the
  Laplacian with Dirichlet boundary condition on $\bd \Appr$ and Neumann
  boundary condition on $\bd \RestAppr \setminus \bd \Appr$.
  Estimate~\eqref{eq:eigen.count2} can be shown similarly.
\end{proof}

To summarize: Theorem~\ref{thm:dec.princ} ensures that a given spectral gap in
the essential spectrum of the periodic manifold remains invariant under local
perturbations. Therefore, $\mathcal N(\tau,\lambda)$ indeed counts the
discrete eigenvalues in the gap.  Theorem~\ref{thm:ef.conv} shows that
eigenfunctions of the approximating problem converge to eigenfunctions of the
full (perturbed) problem and that multiplicity is conserved.
Theorem~\ref{thm:count.approx} follows and with the aid of
Lemma~\ref{lem:approx.probl} we get rid of the approximating index $n$ in
Estimate~\eqref{eq:count.approx}. Thus we have proven our main result
Theorem~\ref{thm:eigenvalues}.

\section{Manifolds with spectral gaps}
\label{sec:mfd.gaps}
For the convenience of the reader, we cite the results on examples of
manifolds with spectral gaps given in \cite{post:01b} and \cite{yoshitomi:98}.
More details and further references can be found therein.

\subsection*{Conformal periodic manifolds}
Let $\Mper$ be a $\Gamma$-periodic manifold of dimension $d$ with periodic
metric $g$.  Furthermore, suppose that $(\rho_\eps)$, $\eps > 0$, is a family
of smooth periodic functions $\map{\rho_\eps} \Mper {(0,\infty)}$, i.e.,
$\rho_\eps(\gamma x)=\rho_\eps(x)$ for all $\gamma \in \Gamma$ and $x \in
\Mper$. We denote by $\Mepsper$ the manifold $\Mper$ with metric $g_\eps :=
\rho_\eps^2 g$.  Similarly we define $\Meps$ for a period cell $M$. We have
the following theorem:
\begin{theorem}
  \label{thm:conf.gaps}
  Suppose $d \ge 3$ and that $\rho_\eps$ converges pointwise on $M$ to the
  indicator function of a closed set $X \subset \overcirc M$ with smooth
  boundary. Then for each $k \in \N$, the $k$-th eigenvalue $\EW k \Meps$ with
  Dirichlet, Neumann or $\theta$-periodic boundary condition converge to $\EWN
  k X$ as $\eps \to 0$ (uniformly in $\theta$). In particular, for all $n \in
  \N$ there exists $\eps>0$ such that $M=\Meps$ satisfies the gap
  condition~\eqref{eq:gap} for $k = 1,\dots,n$ (provided $\EWN k X$ is a
  simple eigenvalue).
\end{theorem}
The precise assumptions on $\rho_\eps$ and $X$ resp.\ $M$ and the proof of
this theorem in the $\theta$-periodic case can be found in \cite{post:00}
resp.\ \cite[Theorem~1.3]{post:01b}. There we also have presented an example
for $d=2$. The proofs for the Dirichlet resp.\ Neumann case are similar.

\subsection*{Attaching small cylindrical ends}
Let $X$ be a compact Riemannian manifold of dimension $d \ge 2$ with metric
$g$. Let $x_1,x_2 \in X$ be two distinct points. On $X \setminus \{x_1,x_2\}$
we change the metric such that a cylinder of radius $\eps$ and length $\eps/2$
is isometrically embedded at each point $x_i$. Here, a cylinder is a product
of the $(d-1)$-dimensional unit sphere $\Sphere^{d-1}$ and an interval. We
denote the (completion of the) resulting manifold by $\Meps$. Note that $\bd
\Meps$ has two additional components $\bd_1 \Meps$ and $\bd_2 \Meps$ each of
them being isometric to $\Sphere^{d-1}$. A typical example of a period cell in
the case when $X$ is a $2$-dimensional torus is drawn in
Figure~\ref{fig:per-cell} on page~\pageref{fig:per-cell}.

For $\gamma \in \Z$ let $\gamma \Meps$ be a copy of $\Meps$. Identifying
$\bd_2 \gamma \Meps$ with $\bd_1 (\gamma+1) \Meps$ pointwise we obtain a
$\Z$-periodic manifold $\Mepsper$ with period cell $\Meps$. Similarly we can
construct $\Gamma$-periodic manifolds by attaching $2r$ cylindrical ends if
$r$ denotes the number of generators of $\Gamma$.  For a detailed construction
we refer to~\cite{post:01b}.

Note that the period cell $\Meps$ always has smooth boundary $\bd M$, i.e., we
can allow non-compact perturbations in the next sections.

Intuitively, the period cell $\Meps$ is close to the original manifold $X$ if
$\eps$ is small. The next theorem shows that the same is true for the
eigenvalues:
\begin{theorem}
  \label{thm:constr.gaps}
  Assume that the periodic manifold $\Mepsper$ and the period cell $\Meps$ are
  constructed as above. Then for each $k \in \N$, the $k$-th eigenvalue
  $\EW k \Meps$ with Dirichlet, Neumann or $\theta$-periodic boundary
  condition converge to $\EW k X$ as $\eps \to 0$ (uniformly in $\theta$). In
  particular, for all $n \in \N$ there exists $\eps>0$ such that $M=\Meps$
  satisfies the gap condition~\eqref{eq:gap} for $k = 1,\dots,n$ (provided
  $\EW k X$ is a simple eigenvalue).
\end{theorem}
Again, the proof of this theorem and related results can be found
in~\cite[Theorem~1.1]{post:01b}. Note that in Theorem~\ref{thm:conf.gaps} as
well as in Theorem~\ref{thm:constr.gaps} the decoupling of the different
period cells $\gamma \Meps$ is responsible for the gaps.

\subsection*{Periodically curved quantum wave guides}
Here, we present an example given by Yoshitomi \cite{yoshitomi:98}. We
consider a $2$-di\-men\-sio\-nal planar strip $\Mepsper$ obtained by
sliding the normal segment of length $\eps$ along a periodically
curved path $\omega$.  Note that the Dirichlet Laplacian on $\Meps$ is
the Hamiltonian for an electron confined in a quantum wire on a planar
substrate, where the vertical dimension is separated
(cf.~\cite{exner-seba:89}).

Suppose $\eps_0 > 0$. Let $\map \kappa \R {(-1/\eps_0,\infty)}$ be a
smooth and $2\pi$-periodic map (i.e., $\kappa(s+2\pi) = \kappa(s)$
for all $s \in \R$). Then
\begin{displaymath}
  \omega(s):= \bigl( x_{\cos}(s),x_{\sin}(s) \bigr) \qquad \text{with} \qquad
  x_f(s):= \int_0^s f \Bigl( -\int_0^{s'} \kappa(s'') \dd s'' \Bigr) \dd s'
\end{displaymath}
for $f=\cos$ or $f=\sin$ is a curve in $\R^2$ with curvature
$\kappa$. We denote by $\dot \omega^\orth(s):=(-\dot x_{\sin}(s), \dot
x_{\cos}(s))$ the normal unit vector with respect to $\dot
\omega(s)$. We set
\begin{equation}
  \label{eq:qwg}
  \Mepsper := 
  \set{\omega(s)+ u \, \dot \omega^\orth(s)}{s \in \R, 0 \le 0 \le \eps}
\end{equation}
and suppose that
\begin{displaymath}
  \map {\Phi_{\eps_0}} {\R \times [0,\eps_0]} \Mepsper, \qquad
  (s,u) \longmapsto \omega(s)+ u \, \dot \omega^\orth(s)
\end{displaymath}
is a diffeomorphism (for the precise assumptions see~\cite{yoshitomi:98}).
This diffeomorphism allows us to calculate the Dirichlet Laplacian on
$\Mepsper$, $0<\eps<\eps_0$, in coordinates $(s,u)$. We furthermore assume
that $\int_0^{2\pi} \kappa(s) \, \dd s =0$. Then $\omega$ is also
$2\pi$-periodic and $\Mepsper$ is indeed a $\Z$-periodic manifold with a
period cell $\Meps$ given by $\Phi_\eps([0,\eps] \times [0, 2\pi])$. We set
\begin{displaymath}
  K := -\frac {\dd ^2}{\dd s^2} - \frac14 \kappa^2.
\end{displaymath}
In $\Lsqr {[0,2\pi]}$ we regard the $\theta$-periodic realisation of $K$ and
denote the corresponding eigenvalues by $\EWT k K$. Yoshitomi proved
that for all $k \in \N$,
\begin{equation}
  \label{eq:qw.asymp}
  \EWT k \Meps = \frac{\pi^2}{\eps^2} + \EWT k K + O(\eps), \qquad \eps \to 0
\end{equation}
uniformly in $\theta$, where $\EWT k \Meps$ denotes the eigenvalues of
the Laplacian on $\Meps$ with Dirichlet boundary condition at $u=0$
and $u=\eps$, and $\theta$-periodic boundary condition at $s=0$ and
$s=2\pi$. A similar identity holds for Dirichlet resp.\ Neumann
boundary condition at $s=0$ and $s=2\pi$ simultaneously for $\laplacian
\Meps$ and $K$.

Furthermore, Yoshitomi proved that if $\kappa \ne 0$ there exists $k
\in \N$ such that the $k$-th band and the $(k+1)$-st band of
$\laplacianD \Mepsper$ are disjoint. He uses classical results about
the inverse problem for Hill's equation
(cf.~\cite{garnett-trubowitz:87}) to show that the periodic
one-dimensional operator $K$ in $\Lsqr \R$ has a gap between the
$k$-th and the $(k+1)$-st band. In our paper, we need the stronger
assumption~\eqref{eq:gap} asserted by the following theorem:
\begin{theorem}
  \label{thm:qw.gaps}
  Assume in addition that the curvature $\kappa$ is even, i.e.,
  $\kappa(-s)=\kappa(s)$.  Then for all $k \in \N$ the $k$-th band $B_k(K)$ of
  the periodic operator $K$ in $\Lsqr \R$ is given by
  \begin{equation}
    \label{eq:qw.gaps}
    B_k(K) = [\EWD k K, \EWN k K].
  \end{equation}
  Here $\EWD k K$ resp.\ $\EWN k K$ denotes the operator $K$ in $\Lsqr
  {[0,2\pi]}$ with Dirichlet resp.\ Neumann boundary conditions.  In particular
  if $\kappa \ne 0$ there exists $k \in \N$ such that the gap
  condition~\eqref{eq:gap} is satisfied for the Dirichlet Laplacian on
  $\Mepsper$ provided $\eps$ is small enough.
\end{theorem}
\begin{proof}
  Since $\kappa$ is even, $K$ has an even potential. In particular,
  \eqref{eq:qw.gaps} holds iff the potential is even
  (cf.~\cite{garnett-trubowitz:87}). The rest follows from
  \cite{yoshitomi:98}.
\end{proof}
Yoshitomi also localized the gaps: he proved that the $k$-th gap of $K$ is
open if the $k$-th Fourier coefficient of $\kappa^2$ is not $0$.
\section{Examples of perturbations}
\label{sec:examples}
In this section we discuss some examples of the great variety of possible
perturbations. The only restriction is the knowledge of eigenvalue estimates
of the perturbed and unperturbed problem to calculate the right hand sides
of~\eqref{eq:from.above} and~\eqref{eq:from.below}. Here, we always assume
that $\Mper$ is a periodic manifold with periodic metric $g$ and period cell
$M$ such that the gap condition~\eqref{eq:gap} is fulfilled.

By the Weyl asymptotic distribution of eigenvalues~\eqref{eq:weyl} and
by~\eqref{eq:phase.diff} we expect that an infinite number of eigenvalue
branches comes from above crossing the level $\lambda$ if we increase the
volume of $\Perttau$ to infinity.  Therefore, we expect an infinite number
of parameters $\tau$, such that $\lambda$ is an eigenvalue of $\laplacian
\Mpertau$ by Theorem~\ref{thm:eigenvalues}.

In contrast, if we shrink the volume of $\Perttau$, we would expect only a
finite number of eigenvalues (depending on the volume of $\Pert$). Here, we
think of finitely many eigenvalue branches crossing the level $\lambda$ from
below.

\subsection*{Conformal perturbations}
First, we give examples of \emph{conformal} perturabtions. Suppose
that $\map {\pert \rho \tau} \Mper {(0,\infty)}$, $\tau \ge 0$, is a
family of smooth functions. The perturbed metric is given by $\pert g
\tau = \pert \rho \tau ^2 g$. Clearly, if the family $(\pert \rho
\tau)$ satisfies
\begin{align}
  \label{eq:conf.unperturbed}
  \pert \rho 0(x) = 1,   & \qquad x \in \Mper \\
  \label{eq:conf.met.cont}
  \norm[C^1] {\pert \rho \tau - \pert \rho {\tau_0}} \to 0
  & \qquad\text{on $\Mper$ as $\tau \to \tau_0$,}\\
  \label{eq:conf.pert.compact}
  \pert \rho \tau(x)= 1, & \qquad x \in \Rest 
\intertext{for all $\tau_0 \ge
    0$ and sufficiently large $m \in \N$ then $(\pert g \tau)$ satisfies
    Conditions~\eqref{eq:unperturbed}, \eqref{eq:met.cont}
    and~\eqref{eq:pert.compact}. Here, the $C^1$-norm of functions is defined
    in the usual way. Furthermore, if $\bd M$ is smooth we can also allow
    non-compact conformal perturbations which are small outside $\Pert$,
    i.e.,}
  \label{eq:conf.met.unif.rest} \tag{6.3'}
  \sup_{\tau \ge 0} \norm[C^1] {\pert \rho \tau - 1} \to 0 
  & \qquad\text{on $\Rest$ as $m \to \infty$.}
\end{align}
If there is a non-compact perturbation we fix $m$ large enough such that
Equation~\eqref{eq:from.above} resp.~\eqref{eq:from.below} holds.

Furthermore, we assume that the conformal factor is constant on some compact
submanifold $K \subset \Pert$ with non-empty interior and piecewise smooth
boundary, i.e.,
\begin{equation}
  \label{eq:conf.const}
  \pert \rho \tau(x)=\pert c \tau, \qquad x \in K
\end{equation}
where $\pert c \tau>0$ is a constant. Again, $\pert K \tau$ denotes the
manifold $K$ with metric $\pert \rho \tau^2 g = \pert c \tau ^2 g$.

\paragraph{Eigenvalues coming from above.}
In our first example we blow up a subset of $\Pert$ by conformal factors:
\begin{proposition}
  \label{prop:conf.blow}
  Suppose that $\lambda \in I_k$ lies between the $k$-th and the $(k+1)$-st
  band and that $\pert c \tau \to \infty$ as $\tau \to \infty$. Then
  \begin{displaymath}
    \mathcal N(\tau,\lambda) \ge
    \frac {\omega_d}{(2\pi)^d} \lambda^{\frac d 2} \vol (\pert K \tau) - km -
      \delta(\tau)
  \end{displaymath}
  where $\delta(\tau) \searrow 0$ as $\tau \to \infty$. In particular,
  $\mathcal N(\tau,\lambda) \to \infty$ as $\tau \to \infty$, i.e., there
  exist an infinite number of parameters $\tau$ such that $\lambda$ is an
  eigenvalue of $\laplacian \Mpertau$.
\end{proposition}

\begin{proof}
  Applying the Min-max principle we conclude
  \begin{displaymath}
    \EWD j \Perttau \le
    \EWD j {\pert K \tau} =
    \pert c \tau ^{-2} \EWD j K \to 0
  \end{displaymath}
  as $\tau \to \infty$ since $\EWD j K > 0$ for all $j \in \N$.  Therefore all
  eigenvalue branches $\tau \mapsto \EWD j \Perttau$, $j \in \N$, cross the
  level $\lambda$ from above (at least once).  Furthermore, there exists a
  positive function $\delta(\tau) \to 0$ as $\tau \to \infty$ such that
  \begin{displaymath}
    \dimEWD \Perttau \ge
    \dimEWD {\pert K \tau}  =
    \dimEWD [\pert c \tau^2 \lambda] K \ge
    \frac {\omega_d}{(2\pi)^d} \lambda^{\frac d 2} \pert c \tau^d \vol (K) -
      \delta(\tau)
  \end{displaymath}
  by the Weyl asymptotic distribution~\eqref{eq:weyl}. The result follows from
  Theorem~\ref{thm:eigenvalues} and Lemma~\ref{lem:common.gap}.
\end{proof}

\paragraph{Eigenvalues coming from below.}
Next we shrink the manifold $\Pert$ on $K$:
\begin{proposition}
  \label{prop:conf.shrink}
  Suppose that $\lambda \in I_k$ and that $\pert c \tau \to 0$ as $\tau \to
  \infty$. Then we have
  \begin{displaymath}
    \mathcal N(\tau,\lambda) \ge k m - 1
  \end{displaymath}
  for $\tau$ sufficiently large. In particular, there exist at least a finite
  number of parameters $\tau$ such that $\lambda$ is an eigenvalue of
  $\laplacian \Mpertau$.
\end{proposition}
\begin{proof}
  The Min-max principle yields
  \begin{displaymath}
    \EWN j \Perttau \ge
    \EWN j {\pert K \tau} =
    \pert c \tau ^{-2} \EWN j K \to \infty
  \end{displaymath}
  as $\tau \to \infty$ for all $j \in \N$, except for the first Neumann
  eigenvalue $\EWN 1 K = 0$.  Therefore the eigenvalue branches $\tau \mapsto
  \EWN j \Perttau$, $j \in \N \setminus \{1\}$ with $\EWN j {\pert \Pert 0} <
  \lambda$, cross the level $\lambda$ from below (at least once).
  Lemma~\ref{lem:common.gap} yields that there are exactly $k m -1$ such
  eigenvalue branches.  Again we apply Theorem~\ref{thm:eigenvalues}.
\end{proof}

\begin{remark}
  \label{rem:mono.conf}
  If $d=2$ one has monotonically decreasing (increasing) eigenvalue branches
  $\tau \mapsto \EWN j \Perttau$ if the conformal factor $\pert \rho \tau$ is
  monotonically increasing (decreasing). Here, the conformal factor only
  occurs in the norm, not in the quadratic form. The monotonicity follows from
  the Min-max principle. This monotonicity can be a first step in the proof
  of an upper bound on the counting function $\mathcal N(\tau,\lambda)$ (see
  the introduction).
\end{remark}

Finally, let us note the following theorem combining results of this section
and of~\cite{post:01b} (cf.~Theorem~\ref{thm:conf.gaps}):
\begin{theorem}
  \label{thm:conformal}
  Let $\Mper$ be a periodic Riemannian manifold of dimension $d \ge 3$ with
  metric $g$. For any $n \in \N$ there exists a periodic metric $g_n$
  conformal to $g$ such that the corresponding Laplacian has at least $n$
  spectral gaps. Furthermore, there exist (non-periodic) metrics on $\Mper$
  conformal to $g$ such that the corresponding Laplacian has an eigenvalue in
  a gap of the essential spectrum.
\end{theorem}

\subsection*{Diffeomorphic perturbations}
\paragraph{Eigenvalues from above.}
Here, we show how perturbations $\tMpertau$ diffeomorphic to $\Mper$ can be
reduced to the case already treated. We think of $\tMpertau$ as being a smooth
deformation of $\Mper$.  For example, consider a quantum wave guide
$\Mper=\Mepsper$ as defined in~\eqref{eq:qwg}. We assume that the gap
condition~\eqref{eq:gap} holds for some fixed $\eps>0$. Let the perturbation
of $\Mper$ be given by cutting the strip at $s=0$ and inserting a ``bubble''
$\pert K \tau$ which blows up as $\tau$ increases.

To be more precise, suppose that $\pert K \tau$ is a simply connected closed
subset of $\R^2$ such that $\pert K \tau$
\begin{itemize}
\item is a rectangle of length $\tau$ and width $\eps$ for small $\tau$, in
  particular, $\pert K 0$ has empty interior;
\item contains a circle of radius $\tau$ or, alternatively, a rectangle of
  length $\tau$ and width $\eps$ for all $\tau$ large enough;
\item has two ends $\pert A {\tau,1}$ and $\pert A {\tau,2}$, each of them
  being isometric to a rectangle of width $\eps$;
\item depends continuously on $\tau$, i.e., the two boundary curves of $\bd
  \pert K \tau \setminus (\pert A {\tau,1} \cup \pert A {\tau,2})$ are smooth
  and depend continuously on $\tau$ and the curve parameter. 
\end{itemize}
\begin{figure}[h]
  \begin{center}
  \begin{picture}(0,0)%
    \includegraphics{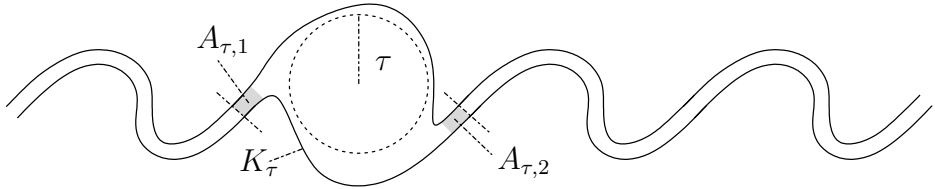}%
  \end{picture}%
  \setlength{\unitlength}{4144sp}
  \begin{picture}(5563,1112)(310,-413)
    \put(2476,299){$\tau$}
    \put(3208,-295){$\pert A {\tau,2}$}
    \put(1663,-310){$\pert K \tau$}
    \put(1408,425){$\pert A {\tau,1}$}
  \end{picture}
    \caption{A quantum wave guide perturbed by inserting an increasing bubble.}
    \label{fig:qwg-pert}
  \end{center}
\end{figure}
We denote the intersected quantum wave guide together with the
inserted bubble by $\tMpertau$ (see Figure~\ref{fig:qwg-pert}). Note
that there exists a diffeomorphism $\map {\pert \Phi \tau} \Mper
\tMpertau$ such that $\pert \Phi \tau(x)=x$ outside a compact set
(with the obvious identification of the two ends of the unperturbed
and perturbed quantum wave guide).

The Euclidean metric $g$ on $\tMpertau$ is pulled back on $\Mper$ via $\pert
\Phi \tau$, i.e., $\pert g \tau := \pert \Phi \tau^* g$. Note that
$(\tau,x) \mapsto \pert g \tau(x)$ resp.\ $(\tau,x) \mapsto \partial_i \pert g
\tau(x)$ are uniformly continuous maps on $[0,\tau_0] \times \Mper$ for all
$\tau_0>0$ since $\pert g \tau = g$ outside a compact set and since $g$ is
periodic, thus uniformly continuous. Therefore, it is easy to check that the
Assumptions~\eqref{eq:unperturbed} to~\eqref{eq:pert.compact} are satisfied.

\begin{proposition}
\label{prop:qw.blow}
Suppose that $\lambda \in I_k$ and that $\tMpertau$ is the perturbed manifold
$\Mper$ obtained by the above construction. Then we have $\mathcal
N(\tau,\lambda) \to \infty$ as $\tau \to \infty$.
\end{proposition}

\begin{proof}
  The proof is similar to the proof of Proposition~\ref{prop:conf.blow}. Note
  that in the case of an inserted rectangle of length $\tau$, we have $\EWD k
  {\pert K \tau} \le (\pi k)^2/\tau^2$ for all $\tau \ge \tau_0$ where
  $\tau_0$ depends on $k$.
\end{proof}

\begin{remark}
  \label{rem:mono.diff}
  Note that the eigenvalue branches $\tau \mapsto \EWN j {\pert K \tau}$ are
  monotonically decreasing by the Min-max principle at least in the case when
  $\pert K \tau$ is isometric to a rectangle of length $\tau$ and width
  $\eps$. This monotonicity can be a first step in the proof of an upper bound
  on the counting function $\mathcal N(\tau,\lambda)$ (see the introduction).
\end{remark}

\section{Topological perturbations}
\label{sec:non.diffeo}
\paragraph{Eigenvalues from below.}
In this final section we show how to deal with certain non-homeomorphic
perturbations. We allow more general examples of perturbed Riemannian
manifolds $\tMpertau$ diffeomorphic to $\MperP := \Mper \setminus H$ where $H$
consists of discrete points. We think of $\tMpertau$ being a smooth
deformation of $\MperP$. Note that the substraction of a discrete set of
points has no influence on the Laplacian as a self-adjoint operator as we will
see now.

Suppose that $X$ is a complete Riemannian manifold of dimension $d \ge 2$.
Let $H \subset X \setminus \bd X$ be a discrete set of points. On $\pointed X
:= X \setminus H$ we define the Laplacian with Dirichlet resp.\ Neumann
boundary condition on $\bd X$ as in Section~\ref{sec:prelim}. At $H$ we always
assume a Dirichlet boundary condition, i.e., we start from the quadratic form
$\check q_X$ defined for smooth functions with support away from $H$. Note
that the Hilbert spaces $\Lsqr X$ and $\Lsqr {\pointed X}$ agree since $H$ has
measure $0$.  Furthermore, the substraction of a discrete set of points $H$
has no effects on the Laplacian either (with any boundary condition):
\begin{theorem}
  \label{thm:spec.equal}
  Suppose that $\bd X=\emptyset$. For every function $u \in \dom q_X$ there
  exists a sequence $(u_n)$ of smooth functions $u_n \in \dom q_X$ with
  support away from $H$ such that $u_n \to u$ in the form norm $(\normsqr[X] u
  + q_X(u))^\frac12$. In particular, $\dom q_{\pointed X} = \dom q_X$ and
  therefore, the corresponding operators agree as operators in $\Lsqr X$.
  
  If $\bd X \ne \emptyset$ the same is true for the quadratic form with
  Dirichlet resp.\ Neumann (or any other) boundary condition on $\bd X$.
  Again, the corresponding Laplacians on the dotted and undotted manifolds
  agree.
\end{theorem}
\begin{proof}
  For a single point see \cite[Lemma~1]{chavel-feldman:78}. Clearly we can
  generalise this result to a discrete set of points $H$. Since we have
  defined the Laplacian via quadratic forms the equality of the operators
  follows.
\end{proof}

We consider the following example: suppose that we bore a hole around each
point of $H$ in the periodic manifold $\Mper$.  For simplicity, we assume that
$H$ consists of $m$ points $\gamma x_0$ where $\gamma \in \Gamma^m$ (recall
that $\Gamma^m$ is an exhausting sequence of $\Gamma$ as defined in the
beginning) and $x_0 \in M$ with $\dist (x_0,\bd M) < \eps_0/2$. Here, $M$
denotes a fixed period cell of $\Mper$ for which the gap
condition~\eqref{eq:gap} holds and $\eps_0>0$ is the injectivity radius of
$\Mper$, i.e., the exponential map is defined on all balls of radius smaller
than $\eps_0$.

\begin{figure}[h]
  \begin{center}
    \begin{picture}(0,0)%
    \includegraphics{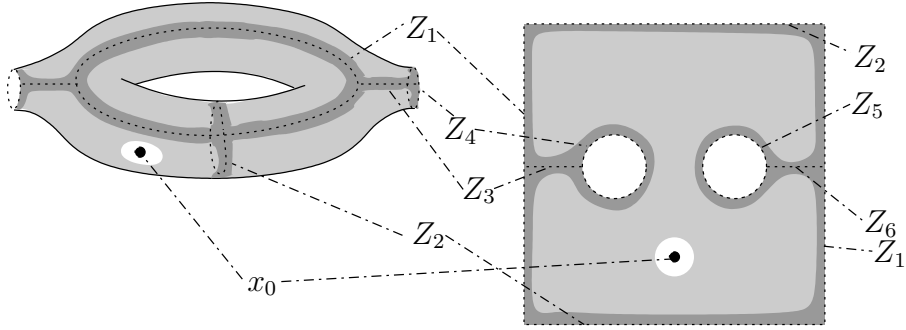}%
    \end{picture}%
    \setlength{\unitlength}{4144sp}%
    \begin{picture}(5198,1951)(49,-1198)
      \put(2441,544){$Z_1$}
      \put(2481,-691){$Z_2$}
      \put(2671,-66){$Z_4$}
      \put(5106, 89){$Z_5$}
      \put(2781,-421){$Z_3$}
      \put(5186,-620){$Z_6$}
      \put(5241,-801){$Z_1$}
      \put(5121,364){$Z_2$}
      \put(1498,-971){$x_0$}
    \end{picture}
    \caption{A period cell with the dotted set $Z$ and its smooth components
      $Z_i$. On the right hand side one has to identify opposite sides of the
      square. Note that the complement of $Z$ in the period cell is always
      homeomorphic to an open ball in $\R^d$. The (light and dark) grey set is
      the perturbed manifold for small $\tau>0$, the dark grey set is the
      perturbed manifold for large $\tau$.}
    \label{fig:per-cell}
  \end{center}
\end{figure}

To define properly the smooth dependence of the boundary on the parameter
$\tau$, we need the following construction: let $\map f \Mper {[0,\infty]}$ be
a continuous map.  We think of $f$ quantifying the level of perturbation:
$f(x)=0$ resp.\ $f(x)=\infty$ means that $x \in \Mper$ is always resp.\ never
affected by the perturbation.  In particular, we assume
\begin{itemize}
\item that $f(x)=\dist(x,\gamma x_0)$ for all $x \in \gamma M$ such that
  $\dist(x, \gamma x_0)<\eps_0/2$, in particular $f(\gamma x_0)=0$ for all
  $\gamma \in \Gamma^m$;
\item that $f(x)=\infty$ for all $x \in \Rest=(\Gamma \setminus \Gamma^m) M$
  and all $x \in \Gamma^m \bd M$, i.e., the union of all translates $\gamma
  \bd M$, $\gamma \in \Gamma^m$;
\item that $f$ is smooth on $\set{x \in \Mper}{0 < f(x)< \infty}$ and that all
  $\tau \in (0,\infty)$ are regular values of $f$, i.e., if $f(x)=\tau$ then
  $\dd f_x \ne 0$;
\item that $Z:=\set{x \in \Pert=\Gamma^m M}{f(x)=\infty}$ is the finite union
  of compact smooth submanifolds $Z_i$ of dimension $d_i<d$ with piecewise
  smooth boundary such that $Z_i \cap Z_j$ is either empty
  or of dimension smaller than $d_i$ and $d_j$ (cf.~Figure~\ref{fig:per-cell});
\item that $\set{x \in \Pert}{f(x)>\tau}$ is contained in the
  $1/\tau$-neighbourhood of $Z$, i.e., $\dist(x,Z)<1/\tau$ for all $x \in
  \Pert$ with $f(x)>\tau$.
\end{itemize}
 Let
\begin{align*}
  \widetilde \Mper  & := 
      \set{(\tau,x) \in [0,\infty) \times \Mper} {f(x) > \tau} 
                          \qquad \text{resp.}\\
  Y             & := \set{(\tau,x) \in [0,\infty) \times \Mper} {f(x) = \tau}
\end{align*}
be the fibred manifold with fibres
\begin{align*}
   \tMpertau    & := \set{x \in \Mper} {f(x) > \tau}  \qquad \text{resp.}\\
   \pert Y \tau & := \set{x \in \Mper} {f(x) = \tau}.
\end{align*}
We consider these manifolds as submanifolds of the product of the
Riemannian manifolds $[0,\infty)$ and $\Mper$ (with the induced
metrics).  Note that $\Mper \setminus \tMpertau$ is a manifold
diffemorphic to $m$ copies of a closed ball in $\R^d$ with smooth
boundary $\pert Y \tau$ diffeomorphic to $m$ copies of the sphere
$\Sphere^{d-1}$ provided $0 < \tau < \infty$. Furthermore, note that
$Y$ consists of $m$ cones. In a neighbourhood of (each component of)
$Y$ we introduce \emph{normal (or Fermi) coordinates} given by a chart
$\map \varphi U V$ where
\begin{displaymath}
  U = \set{(\tau,x) \in \widetilde \Mper}{0 < \dist(x,Y) < r_0(\tau)}
\end{displaymath}
(see e.g.~\cite[p. 9-59 -- 9-62]{spivak:70a}). Here, the function $r_0$ is
supposed to be smooth. Then the image of the chart is given by
\begin{displaymath}
  V = \set{(s,\tau,y) \in (0,\infty) \times [0,\infty) \times Y \,}
          {\, s < r_0(\tau)}.
\end{displaymath}
In particular, for fixed $\tau$ we have a chart $\map{\pert \varphi
  \tau}{\pert U \tau}{\pert V \tau}$ for the corresponding fibres.

Note that $\widetilde \Mper$ is diffeomorphic to $[0,\infty) \times \MperP$. We construct the diffeomorphism explicitly. Let
\begin{displaymath}
  \map r {[0,\infty) \times [0,\infty)} {[0,\infty)}
\end{displaymath}
be a continuous map (smooth on the interior) with bounded derivatives such that
\begin{displaymath}
  \map{\pert r \tau := r(\tau,\cdot)}{[0,\infty)}{[\tau,\infty)} 
\end{displaymath}
is bijective and
\begin{displaymath}
  \pert r \tau(0) = \tau \qquad \text{and} \qquad
  \pert r \tau(s) = s, \quad s \ge r_0(\tau)/2.
\end{displaymath}
Then
\begin{displaymath}
  \map \Phi {[0,\infty) \times \MperP} {\widetilde \Mper}, \qquad \qquad
  \Phi(\tau,x):= \varphi^{-1}(\pert r \tau(s),\tau,y)
\end{displaymath}
with $(s,y):=\pert \varphi \tau(x)$ is a diffeomorphism. Clearly, the
corresponding maps on the fibres $\map{\pert \Phi \tau} \MperP \tMpertau$
are also diffeomorphisms.  For a technical reason we do not use $\MperP$ but
\begin{displaymath}
  \hMper := \set{x \in \Mper} {\dist(x,H) > \eps_0/2}
\end{displaymath}
as a reference manifold. Note that there exists a diffeomorphism
\begin{displaymath}
  \map \psi \hMper \MperP.
\end{displaymath}
Therefore, $\hMper$ is also diffeomorphic to each fibre $\tMpertau$ of
$\widetilde \Mper$. The technical reason is that we need an analogue to
Theorem~\ref{thm:reg.theory} where we have used regularity theory.  On
$\MperP$ we would have coefficients which cannot be continuously extended onto
$H$. Blowing up the holes and pulling back the metrics avoids this difficulty.
Therefore, we have a diffeomorphism
\begin{displaymath}
  \map \Psi {[0,\infty) \times \hMper} {\widetilde \Mper}, \qquad \qquad
  \Psi(\tau,x):=(\tau,\pert \Phi \tau(\psi(x))).
\end{displaymath}

There is a natural metric on $\widetilde \Mper$, namely $\widetilde h:=\dd
\tau^2 + g$ where $g$ is the (restriction of the) periodic metric on $\Mper$.
Let $h:= \Psi^* \widetilde h$ be the pull-back of the metric $\widetilde h$.
Note that $h$ (and its derivatives) are uniformly continuous maps on
$[0,\tau_0] \times \hMper$.  Therefore, the restriction $\pert g \tau$ of $h$
onto the fibre $\hMper$ satisfies the continuity property~\eqref{eq:met.cont}.
Furthermore, Condition~\eqref{eq:pert.compact} is fulfilled.

Finally, replacing $\Mper$ by $\hMper$ and $\GpM$ by $\GpM \cap \hMper$ with
metric $\psi^*g$ one can verify that all our results of
Sections~\ref{sec:ell.est} and~\ref{sec:eigenvalues} remain true in virtue of
Theorem~\ref{thm:spec.equal}.

Therefore, we have the following result:
\begin{proposition}
  \label{prop:diff.shrink}
  Suppose that $\lambda \in I_k$. Let $(\tMpertau)$ be the family of manifolds
  constructed above by removing closed sets from $\Pert$. Then we have
  $\widetilde {\mathcal N}(\tau,\lambda) \ge k m$ for $\tau$ sufficiently
  large.  Here, $\widetilde{\mathcal N}(\tau,\lambda)$ denotes the counting
  function for the manifold $\tMpertau$ defined as
  in~\eqref{eq:def.count.fct}. In particular, there exist a finite number of
  parameters $\tau$ such that $\lambda$ is an eigenvalue of $\laplacian
  \tMpertau$.
\end{proposition}
\begin{proof}
  The proof is the same as the proof of Proposition~\ref{prop:conf.shrink}.
  Set $\pert {\widetilde M^m} \tau := \tMpertau \cap \Pert$.  We need the fact
  that $\EWN j {\pert {\widetilde M^m} \tau} \to \infty$ as $\tau \to \infty$
  for all $j \in \N$ when $\vol (\pert {\widetilde M^m} \tau) \to 0$ as $\tau
  \to \infty$. This will be shown in the next lemma.
\end{proof}

\begin{remark}
  Note that we have imposed Neumann boundary condition only at the
  $\tau$-independent component $\bd \Pert = \bd \pert {\widetilde M^m} \tau$ and
  Dirichlet boundary condition at $\pert Y \tau$. If there was only Neumann
  boundary condition the first eigenvalue would be $0$.
\end{remark}

\begin{remark}
  \label{rem:monotonical}
  In this example, the eigenvalue branches $\tau \mapsto \EWN j {\pert
    {\widetilde M^m} \tau}$ are monotonically increasing due to the Min-max
  principle.  This monotonicity can be a first step in the proof of an upper
  bound on the counting function $\mathcal N(\tau,\lambda)$ (see the
  introduction).
\end{remark}

In the next lemma we show that the eigenvalues grow up to infinity
when shrinking the volume of a manifold to $0$.
\begin{lemma}
  \label{lem:ew.mfd.shrink}
  Suppose that $K$ is a $d$-dimensional manifold with piecewise smooth
  boundary $\bd K$. Denote by $(\pert K \tau)$ a family of $d$-dimensional
  closed submanifolds of $K$ such that $\vol \pert K \tau \to 0$ as $\tau \to
  \infty$ and such that $\bd \pert K \tau$ depends smoothly on $\tau$ (defined
  in the same way as before). Furthermore, assume that $\bd \pert K \tau$
  consists of two disjoint components $\pert Y \tau$ and $\bd K$
  where $\pert Y \tau$ is smooth. Besides, suppose that
  \begin{displaymath}
    \pert K \tau \subset 
    \set{x \in K}{\dist(x,Z) < \frac 1 \tau} := 
    \pert N \tau
  \end{displaymath}
  for $\tau$ large. Here, we assume that $Z$ is the finite union of compact
  smooth submanifolds $Z_i$ of dimension $d_i<d$ with piecewise smooth
  boundary such that $Z_i \cap Z_j$ is either empty or of dimension smaller
  than $d_i$ and $d_j$ (cf.~Figure~\ref{fig:per-cell}).  Finally, suppose that
  $\bd K \subset Z$.
  Then $\EWDN k {\pert K \tau} \to \infty$ as $\tau \to \infty$ for all $k \in
  \N$ where we have imposed Dirichlet boundary conditions on $\pert Y \tau$
  and Neumann boundary condition on $\bd K$.
\end{lemma}
\begin{proof}
  We define the following isoperimetric (or Cheeger's) constant
  \begin{displaymath}
    h_{\bd K}(\pert K \tau) := 
    \inf_{\pert \Omega \tau} 
      \frac{\vol_{d-1}(\bd \pert \Omega \tau \setminus \bd K)}
           {\vol_d   (\pert \Omega \tau)}
  \end{displaymath}
  where $\pert \Omega \tau$ ranges over all open subsets of $\pert K \tau$
  such that $\bd \pert \Omega \tau \setminus \bd K$ is smooth and $\bd \pert
  \Omega \tau \cap \pert Y \tau = \emptyset$. The subscript $\bd K$ indicates
  the subset of $\bd \pert K \tau$ where we have imposed Neumann boundary
  conditions. When measuring the surface we exclude the Neumann boundary part
  $\bd K$.  If there were no Dirichlet boundary condition (i.e., $\bd \pert K
  \tau=\bd K$) we could choose $\pert \Omega \tau=\pert K \tau$.  Therefore,
  $\bd \pert \Omega \tau \setminus \bd K = \emptyset$, i.e., $h_{\bd K}(\pert
  K \tau)=0$.
  
  In the same way as in \cite[Chapter~IV]{chavel:84} we can prove the
  following estimate originally due to Cheeger:
  \begin{displaymath}
    \EWDN 1 {\pert K \tau} \ge \frac14 h_{\bd K}(\pert K \tau)^2.
  \end{displaymath}
  \begin{figure}[h]
    \begin{center}
      \begin{picture}(0,0)%
    \includegraphics{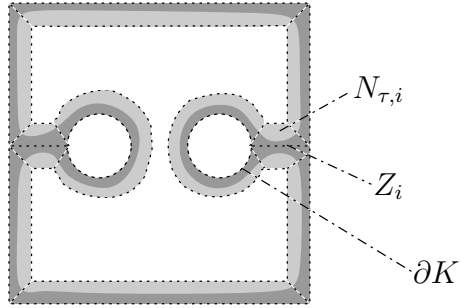}%
      \end{picture}%
      \setlength{\unitlength}{4144sp}%
      \begin{picture}(2454,1824)(934,-1198)
        \put(3376,-1051){$\bd K$}
        \put(3016, 59){$\pert N {\tau,i}$}
        \put(3139,-521){$\pert Z i$}
      \end{picture}
      \caption{The decomposition of the tubular neighbourhood of $Z$. Again,
        we have to identify opposite sides of the square. Here, $K$ is the
        torus with two open balls cut out, $\pert K \tau$ is marked in dark
        grey and the tubular neighbourhood $\pert N \tau$ of radius $1/\tau$
        is marked in light and dark grey. Note that near $Z_i$,
        we have to decompose the tubular neighbourhood once more into two
        sets.}
      \label{fig:decomp}
    \end{center}
  \end{figure}
  It remains to show that $h_{\bd K}(\pert K \tau) \to \infty$ as $\tau \to
  \infty$.  To this end let $\pert \Omega \tau$ be one of the sets taken in
  the definition of $h_{\bd K}(\pert K \tau)$. Since the set $Z$ is the finite
  union of smooth submanifolds $Z_i$ we decompose the (subset of a) tubular
  neighbourhood $\pert N \tau$ of $Z$ into a finite number of disjoint open
  sets $\pert N {\tau,i}$ such that $Z_i=\pert N {\tau,i} \cap Z$ and such
  that $\pert N \tau \setminus \bigcup_i \pert N {\tau,i}$ has $d$-dimensional
  volume $0$.  If $d_i=d-1$ the set $\pert N {\tau,i}$ could lie on both sides
  of $Z_i$ (cf.\ Figure~\ref{fig:decomp}). In this case we decompose $\pert N
  {\tau,i}$ into its two components of $\pert N {\tau,i} \setminus Z_i$ (again
  denoted by $\pert N {\tau,i}$).
  
  We set $\pert A {\tau,i}:= (\bd \pert \Omega \tau \cap \pert N {\tau,i})$
  and $\pert \Omega {\tau,i}:= \pert \Omega \tau \cap \pert N {\tau,i}$. With
  regard to the next lemma, we have
  \begin{displaymath}
    \vol_{d-1}(\bd \pert \Omega \tau \setminus \bd K) \ge
    \sum_i \vol_{d-1}(\pert A {\tau,i} ) \ge
    c \, \tau \sum_i \vol_d (\pert \Omega {\tau,i}) =
    c \, \tau \vol_d (\pert \Omega \tau)
  \end{displaymath}
  and we are done.
\end{proof}

We still need the following final technical lemma which roughly says that the
``local'' isoperimetric constant in a small strip tends to infinity when the
width of the strip tends to $0$.
\begin{lemma}
  \label{lem:loc.isoper}
  With the notation from above we have
  \begin{displaymath}
     \vol_{d-1}(\pert A {\tau,i} ) \ge
     c \, \tau \vol_d (\pert \Omega {\tau,i})
  \end{displaymath}
  where the constant $c$ only depends on the metric near $Z$.
\end{lemma}
\begin{proof}
  Since $Z_i$ is a smooth submanifold of $K$ we can introduce normal
  coordinates $(y,z)$ on the tubular neighbourhood $\pert N {\tau,i}$ of $Z_i$
  provided $\tau$ is large enough, i.e., $\pert N {\tau,i}$ is diffeomorphic
  to $\pert B \tau \times (Z_i \setminus \bd Z_i)$ where $\pert B \tau:=\set{y
    \in \R^{d-d_i}}{|y| < 1/\tau}$ (resp.\ $\pert B \tau:=(0,1/\tau)$ if
  $d_i=d-1$), see e.g.~\cite[p. 9-59 -- 9-62]{spivak:70a}.  Since $Z_i$ is
  compact and since the metric on $\pert N {\tau,i}$ depends continuously on
  the coordinates and smoothly on $\tau$ it suffices to prove the assertion in
  the case when the metric (in normal coordinates) has product structure $\dd
  y^2 + h$ where $h$ is the metric on $Z_i$.
  
  Denote by $\pert P {\tau,i}$ the image of $\pert A {\tau,i}$ under the
  orthogonal projection onto $Z_i$.  If we parametrize the smooth submanifold
  $\pert A {\tau,i}$ we can show that
  \begin{displaymath}
    \vol_{d-1} (\pert A {\tau,i}) \ge 
    c' \, \tau^{-(d-1-d_i)} \vol_{d_i} (\pert P {\tau,i})
  \end{displaymath}
  where the constant $c'$ depends only on the metric $h$. Finally,
  \begin{displaymath}
     \vol_d (\pert \Omega {\tau,i}) \le
     \vol_d (\pert N {\tau,i}) \le
     c''\, \tau^{-(d-d_i)} \vol_{d_i} (\pert P {\tau,i})
   \end{displaymath}
   where $c''$ depends only on $d-d_i$. Therefore we have finished our proof.
\end{proof}

\section*{Acknowledgements}
  I would like to thank Rainer Hempel for helpful discussions leading to this
  article. Furthermore I would like to thank Wolf Jung for his suggestions
  concerning the proofs of Lemmas~\ref{lem:ew.mfd.shrink}
  and~\ref{lem:loc.isoper}.


\providecommand{\bysame}{\leavevmode\hbox to3em{\hrulefill}\thinspace}

\end{document}